%% file: sample.tex
\documentclass[twoside,11pt]{article}

\usepackage{blindtext}

\usepackage[preprint]{jmlr2e}
\usepackage{amsmath}
\usepackage{amsfonts}
\usepackage{nicefrac}       %
\usepackage{microtype}      %
\usepackage{xcolor} 
\usepackage{subcaption}
\usepackage{algorithm}
\usepackage{algpseudocode}
\usepackage{bbm}
\usepackage{cleveref}

\usepackage{lastpage}
\jmlrheading{}{2025}{1-\pageref{LastPage}}{9/25}{}{21-0000}{Kushagra Gupta, Xinjie Liu, Ross E. Allen, Ufuk Topcu and David Fridovich-Keil}

\ShortHeadings{Second-Order Algorithms for Finding Local Nash Equilibria in Zero-Sum Games}{Gupta, Liu, Allen, Topcu and Fridovich-Keil}
\firstpageno{1}

\begin{document}

\title{Second-Order Algorithms for Finding Local Nash Equilibria in Zero-Sum Games}

\author{\name Kushagra Gupta \email kushagrag@utexas.edu \\
       \addr Department of Electrical and Computer Engineering\\
       The University of Texas at Austin\\
       Austin, TX, 78712, USA
       \AND
       \name Xinjie Liu \email xinjie-liu@utexas.edu \\
       \addr Department of Electrical and Computer Engineering\\
       The University of Texas at Austin\\
       Austin, TX, 78712, USA
       \AND
       \name Ross E. Allen \email ross.allen@ll.mit.edu \\
       \addr MIT Lincoln Laboratory\\
       Lexington, MA, 02420, USA
       \AND
       \name Ufuk Topcu \email utopcu@utexas.edu \\
       \addr Department of
Aerospace Engineering and Oden Institute for Computational
Engineering and Sciences\\
       The University of Texas at Austin\\
       Austin, TX 78712, USA
       \AND
       \name David Fridovich-Keil \email dfk@utexas.edu \\
       \addr Department of
Aerospace Engineering and Oden Institute for Computational
Engineering and Sciences\\
       The University of Texas at Austin\\
       Austin, TX 78712, USA}

\editor{}

\maketitle

\input{sections/00_abstract}

\input{macros}
\input{sections/01_introduction}
\input{sections/related_work}
\input{sections/02_prelims}
\input{sections/03_methods}

\input{sections/04_Rates}
\input{sections/05_exps}

\input{sections/06_conc}


\acks{
DISTRIBUTION STATEMENT A. Approved for public release. Distribution is unlimited.
This material is based upon work supported by the Under Secretary of Defense for Research and Engineering under Air Force Contract No. FA8702-15-D-0001 or FA8702-25-D-B002. Any opinions, findings, conclusions or recommendations expressed in this material are those of the author(s) and do not necessarily reflect the views of the Under Secretary of Defense for Research and Engineering.
© 2025 Massachusetts Institute of Technology.
Delivered to the U.S. Government with Unlimited Rights, as defined in DFARS Part 252.227-7013 or 7014 (Feb 2014). Notwithstanding any copyright notice, U.S. Government rights in this work are defined by DFARS 252.227-7013 or DFARS 252.227-7014 as detailed above. Use of this work other than as specifically authorized by the U.S. Government may violate any copyrights that exist in this work.

This work also acknowledges funding from the following grants: Synthesis of Strategies for Integrated Mission Planning, Resource Management, and Information Acquisition (ONR N00014-24-1-2797), Attack-Resilient Mission Planning for Swarms (ONR N00014-22-1-2703), DECODE AI (ONR N00014-24-1-2432) and National Science Foundation CAREER award under Grant No. 2336840.
}


\newpage

\appendix
\input{sections/app}









\vskip 0.2in
\bibliography{sample}

\end{document}

%% file: sections/00_abstract.tex
\begin{abstract}
    Zero-sum games arise in a wide variety of problems, including robust optimization and adversarial learning. However, algorithms deployed for finding a local Nash equilibrium in these games often converge to non-Nash stationary points. This highlights a key challenge: for any algorithm, the stability properties of its underlying dynamical system can cause non-Nash points to be potential attractors. To overcome this challenge, algorithms must account for subtleties involving the curvatures of players' costs. To this end, we leverage dynamical system theory and develop a second-order algorithm for finding a local Nash equilibrium in the smooth, possibly nonconvex-nonconcave, zero-sum game setting. First, we prove that this novel method guarantees convergence to only local Nash equilibria with an asymptotic local \textit{linear} convergence rate. We then interpret a version of this method as a modified Gauss-Newton algorithm with local \textit{superlinear} convergence to the neighborhood of a point that satisfies first-order local Nash equilibrium conditions. In comparison, current related state-of-the-art methods with similar guarantees do not offer convergence rates in the nonconvex-nonconcave setting. Furthermore, we show that this approach naturally generalizes to settings with convex and potentially coupled constraints while retaining earlier guarantees of convergence to only local (generalized) Nash equilibria. Code for our experiments can be found at \url{https://github.com/CLeARoboticsLab/ZeroSumGameSolve.jl}.
\end{abstract}

%% file: macros.tex
\newcommand{\x}{\mathbf{x}}
\newcommand{\y}{\mathbf{y}}
\newcommand{\z}{\mathbf{z}}
\newcommand{\zk}{\z_k}
\newcommand{\zkone}{\z_{k+1}}
\newcommand{\om}{\omega(\z)}
\newtheorem{assumption}{Assumption}
\newcommand{\intg}{\mathrm{int}~\mathcal{G}}
\newcommand{\boundg}{\partial\mathcal{G}}
\newcommand{\gset}{\mathcal{G}}
\newcommand{\gd}{g_d}
\newcommand{\second}{\texttt{SecOND}}
\newcommand{\ellfunc}[1]{\frac{1}{2}\|\omega(#1)\|^2}
\newcommand{\kushy}[1]{\textcolor{red}{#1}}

%% file: sections/01_introduction.tex
\section{Introduction}\label{section: intro}

We consider the setting of smooth, deterministic two-player zero-sum games of the form 
\begin{equation}\label{eq: min x f max y f}\tag{Game 1}
    \mathrm{Player\,1:\,} \min_{\x} f(\x, \y) \,\, \quad \quad \quad \,\,\mathrm{Player\,2:\, } \max_{\y} f(\x, \y) \quad \quad (\x, \y)\in\mathcal{G},
\end{equation} where $f$ can be nonconvex-nonconcave with respect to $\x\in\mathbb{R}^n$ and $\y\in\mathbb{R}^m$, respectively. In the unconstrained setting, i.e., when $\mathcal{G}$ is $(\mathbb{R}^n, \mathbb{R}^m)$, we seek to find a local Nash equilibrium. For the constrained setting, we will assume that $\mathcal{G}$ is convex and seek a local generalized Nash equilibrium. \par
Mathematical games are commonly studied in decision-making scenarios involving multiple agents in control theory \citep{isaacs1999differential}, economics \citep{roth2002economist, rubinstein1982perfect}, and computer science \citep{roughgarden2010algorithmic}. In particular, several problems of interest have a natural zero-sum game formulation, such as training generative adversarial networks \citep{goodfellow2014generative}, pursuit-evasion scenarios \citep{isaacs1999differential}, and robust optimization \citep{ben2009robust}. \par
Several recent efforts \citep{jin2020local, fiez2020implicit, Wang*2020On, chinchilla2023newton} consider a closely related minimax variant of \ref{eq: min x f max y f}, \(\min_{\x} \max_{\y} f(\x,\y)\); however, (local) \emph{minimax} solutions can differ from (local) Nash equilibria in general nonconvex-nonconcave settings. 
This difference arises from the \emph{order} of agent interactions. At a Nash solution of \ref{eq: min x f max y f}, players controlling $\x$ and $\y$ act \emph{simultaneously}. In contrast, minimax points correspond to Stackelberg equilibria and assert a \emph{sequential} order of play: $\x$ acts first, then $\y$ follows. We highlight this fact to point out that under the assumptions of \ref{eq: min x f max y f}, the set of all local Nash points is a subset of the set of all local Stackelberg points \citep{mazumdar2020gradient, ratliff2016characterization}. In particular, local Nash and Stackelberg points have the same first-order conditions but different second-order conditions.\par
The success of first-order gradient methods for single-agent learning problems made gradient descent ascent (GDA), the 
two-player zero-sum analog of gradient descent, a natural starting point for solving \ref{eq: min x f max y f}. The GDA algorithm tries to find a critical point of \(f\), i.e., where \(\nabla f_{\x,\y} = \mathbf{0}\). 
GDA is known to get trapped in limit cycles even in the most straightforward convex-concave setting, and several works have tried to modify the gradient dynamics by including second-order information to avoid this entrapment and direct the solution towards a stationary point of the dynamics \citep{benaim1999mixed, daskalakis2017training, hommes2012multiple, mertikopoulos2018cycles, mescheder2017numerics, gidel2019negative}. However, outside of the convex-concave setting, these methods can converge to critical points that are \emph{not} Nash equilibria. This behavior is due to the particular structure of second-order derivatives of \(f\) with respect to \(\x\) and \(\y\), and while they do not arise in the single-agent settings, they are widely documented in the 
two-agent zero-sum game setting \citep{ratliff2016characterization,balduzzi2018mechanics,mazumdar2020gradient}.\par
To guarantee that an algorithm only converges to local Nash equilibria, the algorithm's 
dynamics must not have any non-Nash locally stable equilibrium points. To the best of our knowledge, only two previous works, local symplectic surgery (LSS) \citep{mazumdar2019finding} and curvature exploitation for the saddle point problem (CESP) \citep{adolphs2019local}, have such guarantees for the unconstrained nonconvex-nonconcave version of \ref{eq: min x f max y f}. However, neither of these methods provides any convergence rate analysis for the nonconvex-nonconcave regime. Further, these works do not discuss the constrained setting of \ref{eq: min x f max y f}. A variety of Bregman proximal algorithms do find local min-max points in constrained, nonconvex-nonconcave settings with at best linear rates of convergence; however, they operate under the restrictive, blanket assumption that every critical point of \(f\) is a local Nash equilibrium \citep{Azizian2024}, which is not generally true in nonconvex-nonconcave settings.

In this paper, we introduce second-order algorithms to solve \ref{eq: min x f max y f}. We highlight our specific contributions below:
\begin{enumerate}
    \item We introduce \textbf{D}iscrete-time \textbf{N}ash \textbf{D}ynamics (\texttt{DND}), a discrete-time dynamical system that provably converges to only local Nash equilibria of the unconstrained version of \ref{eq: min x f max y f} with a linear asymptotic local convergence rate for nonconvex-nonconcave games. In contrast, previous related works with similar guarantees do not provide any convergence rates for the same nonconvex-nonconcave setting.
    \item We modify this dynamical system and construct an algorithm, \textbf{Sec}ond \textbf{O}rder \textbf{N}ash \textbf{D}ynamics (\texttt{SecOND}), which can converge superlinearly to the neighborhood of a point that satisfies first-order local Nash conditions.
    \item We show that in structured settings, \texttt{SecOND} exhibits nonasymptotic last-iterate convergence rates that are competetive to previous works with similar guarantees---specifically---linear convergence in bilinear games; and global linear and local quadratic convergence in a broader class of convex-concave games.
    \item We discuss the constrained setting of \ref{eq: min x f max y f}, where \(\mathcal{G}\) is a convex set. In this case, we use Euclidean projections to modify \texttt{DND} and develop an algorithm, \textbf{Se}cond-order \textbf{Co}nstrained \textbf{N}ash \textbf{D}ynamics (\texttt{SeCoND}), which finds a local generalized Nash Equilibrium point. In contrast, previous work either does not consider this constrained setting and/or is restricted to the convex-concave case.
\end{enumerate} 

%% file: sections/related_work.tex
\section{Related Work}
\paragraph{The convex-concave setting.} The study of zero-sum games is classical. Seminal work for zero-sum games considered the bilinear case, with strategies constrained to lie in the probability simplex, and established that the minimax values of such games was equal to the corresponding maximin values, i.e., $\min_{\x\in\Delta^{n-1}}\max_{\y\in\Delta^{m-1}}\x^\top A \y=\max_{\y\in\Delta^{m-1}}\min_{\x\in\Delta^{n-1}}\x^\top A \y,$ $ ~A\in\mathbb{R}^{n\times m}$ \citep{v1928theorie}. This result was later generalized to hold for any convex-concave function $f$ and strategies $\x,\y$ lying in compact convex sets \citep{sion1958general}. These foundational works have 
given rise to
a plethora of gradient-based optimization algorithms that seek (global) Nash equilibria, or saddle points, in convex-concave zero-sum games---with single/double-timescale GDA, proximal point, extra-gradient and optimistic gradient algorithms being the subject of extensive convergence analysis in this setting \citep{rockafellar1976monotone, liang2019interaction, mokhtari2020unified, azizian2020tight, facchinei2003finite, tseng1995linear, gidel2018variational, daskalakis2017training}.
\paragraph{The nonconvex-nonconcave setting.} In the nonconvex-nonconcave setting, gradient-based methods often fail to find Nash equilibria, due to iterates either (i) collapsing to oscillatory cycles, or (ii) converging to non-Nash points. Several works focused on incorporating second order information into gradient-based algorithms to reduce cycling behavior \citep{benaim1999mixed, hommes2012multiple, mescheder2017numerics, daskalakis2017training, mertikopoulos2018cycles, gidel2019negative}. However, it was shown that these dynamics can still converge to non-Nash points in nonconvex-nonconcave settings \citep{mazumdar2019finding}. Many works avoid these problems by considering structured variants of \ref{eq: min x f max y f} that satisfy a variety of regularity assumptions---such as the Minty condition \citep{mertikopoulos2018optimistic, diakonikolas2021efficient}, second-order sufficiency \citep{Azizian2024}, or comonotonicity \citep{cai2024accelerated}. However, outside of these structural assumptions, in more general nonconvex-nonconcave settings, a fixed point need not necessarily be a local Nash equilibrium. As mentioned in \Cref{section: intro}, only two previous works, CESP \citep{adolphs2019local} and LSS \citep{mazumdar2019finding} guarantee convergence to only (strict) local Nash equilibria, with both introducing algorithms employing second-order derivative information. However, neither works establish convergence rates for the nonconvex-nonconcave setting, or consider the constrained version of \ref{eq: min x f max y f}.

Other related directions of work consider different notions of equilibrium in the zero-sum nonconvex-nonconcave setting. One such notion is the local (Stackelberg) minimax equilibrium \citep{jin2020local}---which encodes an order of play, with the minimizing player acting first. For such a setting, various algorithms with convergence to only local minimax equilibria have been introduced, including constructions of two timescale GDA \citep{fiez2021local} and second-order algorithms \citep{fiez2020implicit, Wang*2020On}. Another notion is a relaxation of the local Nash equilibrium---\emph{approximate} local minmax equilibrium, which corresponds to approximations of fixed points of the projected GDA dynamics for \ref{eq: min x f max y f} \citep{daskalakis2021complexity}. Recently, algorithms have been developed for finding such approximate local minmax equilibrium under hypercube constraints \citep{daskalakis2023stay} and more general hyperball constraints \citep{attias2025fixed}. However, we reiterate that both notions differ from the concept of local Nash equilibrium that we are interested in, and that in nonconvex-nonconcave settings, there can exist points which are both local Stackelberg and approximate local minmax equilibria, but not local Nash equilibria \citep{mazumdar2020gradient, jin2020local, daskalakis2021complexity}. 
\paragraph{Other settings.} Several other variants of \ref{eq: min x f max y f} have been considered in the literature. 
For example, the case of nonconvex-concave $f$ is widely studied \citep{grnarova2017online, thekumparampil2019efficient, lin2020gradient, boct2023alternating, lin2025two}---this case naturally arises when the maximizer can find a global best response to the minimizer's strategy, but the minimizer can only find local best responses to the maximizer.
However, due to the differing assumptions on the maximizing player, results from this setting do not always necessarily apply to our setting (and vice-versa).

%% file: sections/02_prelims.tex
\section{Preliminaries}
Throughout this paper, \(\x \in \mathbb{R}^n, \y \in \mathbb{R}^m,\) and \(\z = (\x^\top, \y^\top)^\top \in \mathbb{R}^{n+m}\).
\subsection{Game-theoretic Concepts}
\begin{definition} \textnormal{\textbf{(Strict local Nash equilibrium)}} A strategy \(({\x^*}, {\y^*}) \in \mathbb{R}^n\times\mathbb{R}^m\) is a strict local Nash equilibrium of \ref{eq: min x f max y f}, if
\begin{equation}
    f(\x^*, \y) < f(\x^*, \y^*) < f(\x, \y^*) ,
\end{equation}
for all \(\x\) and \(\y\) in feasible neighborhoods of \({\x^*}\) and \({\y^*}\) respectively.
\end{definition}
Under the smoothness assumption of \ref{eq: min x f max y f}, defining first-order and second-order equilibrium conditions can help identify whether a point is a local Nash equilibrium \citep{ratliff2016characterization}. For the unconstrained setting, any point that satisfies the conditions below is said to be a differential Nash equilibrium and is guaranteed to be a strict local Nash equilibrium.
\begin{definition} \label{definition: DNE}
    \textnormal{\textbf{(Sufficient conditions for strict local Nash equilibrium)}} A strategy \(({\x^*}, {\y^*}) \in \mathbb{R}^n\times\mathbb{R}^m\) is a differential Nash equilibrium (and thus, a strict local Nash equilibrium) of \ref{eq: min x f max y f} when \(\mathcal{X}\) is \(\mathbb{R}^n\) and \(\mathcal{Y}\) is \(\mathbb{R}^m\), if
    \begin{equation} \label{eq: strict LNE conditions}
        \begin{split}
            \nabla_{\x} f(\x^*, \y^*) = 0, & \quad \nabla_{\y} f(\x^*, \y^*) = 0 \\
            \nabla^2_{\x\x} f(\x^*, \y^*) \succ 0, & \quad \nabla^2_{\y\y} f(\x^*, \y^*) \prec 0.
        \end{split}
    \end{equation}
\end{definition}
We now discuss the constrained version of \ref{eq: min x f max y f}. This paper allows the constrained setting to have coupled constraints. In the presence of coupling, the Nash equilibrium sought is a generalized Nash equilibrium.
\begin{definition}
    \textnormal{\textbf{(Local generalized Nash equilibrium)}} Assume the set \(\mathcal{G}\) is convex. A strategy \(({\x^*}, {\y^*}) \in \mathcal{G}\) is a local generalized Nash equilibrium of \ref{eq: min x f max y f} if
    \begin{equation}
        \begin{split}
            f(\x^*, \y^*) & \leq f(\x, \y^*) \, \forall \, (\x, \y^*) \in \mathcal{G}\textnormal{ in a neighborhood around }(\x^*, \y^*) \\
            f(\x^*, \y^*) & \geq f(\x^*, \y) \, \forall \, (\x^*, \y) \in \mathcal{G}\textnormal{ in a neighborhood around }(\x^*, \y^*).
        \end{split}
    \end{equation}
\end{definition}
The optimality conditions of generalized Nash equilibria in the above-mentioned settings are well studied \citep{facchinei2010generalized, facchinei2010penalty}. Though a standard treatment would involve defining the Karush-Kuhn-Tucker conditions for \ref{eq: min x f max y f}, for our purpose, the following conditions are sufficient for a point to be a local generalized Nash equilibrium.
\begin{definition}\label{definition: GNE}
    \textnormal{\textbf{(Sufficient conditions for local generalized Nash equilibrium)}} Assume the set \(\mathcal{G}\) is convex. Let $\boundg$ denote the set of boundary points of $\gset$ and let \(\mathcal{N}(\x, \y)\) denote a neighbourhood around \((\x,\y)\). Then:
    \begin{itemize}
        \item If for a strategy \((\x^*, \y^*) \in \gset\),
        \begin{equation*}
            \begin{split}
            \nabla_{\x} f(\x^*, \y^*) = 0, & \quad \nabla_{\y} f(\x^*, \y^*) = 0 \text{ and} \\
            \nabla^2_{\x\x} f(\x^*, \y^*) \succ 0, & \quad \nabla^2_{\y\y} f(\x^*, \y^*) \prec 0,
        \end{split}
        \end{equation*}
        then \((\x^*, \y^*)\) is a \emph{strict} local generalized Nash equilibrium of \ref{eq: min x f max y f}.
        \item If for a strategy \((\x^*, \y^*) \in \boundg\)
        \begin{equation*}
            \left(\begin{bmatrix} \x \\ \y \end{bmatrix} - \begin{bmatrix} \x^* \\ \y^* \end{bmatrix}\right)^\top \begin{bmatrix} \nabla_\x f(\x^*, \y^*) \\ - \nabla_\y f(\x^*, \y^*)\end{bmatrix} > 0 \,\, \forall \, (\x, \y) \in  \mathcal{G}/(\x^*, \y^*) \cap \mathcal{N}(\x^*, \y^*)\,
        \end{equation*}
        then \((\x^*, \y^*)\) is a \emph{strict} local generalized Nash equilibrium of \ref{eq: min x f max y f}. The strictness is lost if the inequality can hold with equality.
    \end{itemize}
\end{definition}
We now describe some concepts from dynamical system theory that determine whether an algorithm can converge to a local Nash equilibrium.
\subsection{A Dynamical Systems Perspective}
We illustrate how considerations of dynamical system theory are naturally motivated in our work through the example of GDA. We define:
\begin{equation}\label{eq: notation omega and J}
    \omega (\z) := \begin{bmatrix}
        \nabla_{\x} f(\x, \y) \\ - \nabla_{\y} f(\x,\y)
    \end{bmatrix}, \quad J(\z) := \nabla_{\z} \omega(\z) = \begin{bmatrix}
        \nabla^2_{\x\x}f(\x, \y) & \nabla^2_{\x\y}f(\x, \y) \\
        - \nabla^2_{\y\x}f(\x, \y) & - \nabla^2_{\y\y}f(\x, \y)
    \end{bmatrix}.
\end{equation}
For some stepsize $\gamma$, the GDA update for \ref{eq: min x f max y f} for any iteration \(k\) can thus be written as
\begin{equation} \label{eq: GDA discrete dynamical sys}
    \z_{k+1} = g_{\mathrm{GDA}}(\z_k):= \z_k - \gamma \omega(\z_k).
\end{equation}
\Cref{eq: GDA discrete dynamical sys} can be viewed as a discrete-time dynamical system. We may also consider the limiting ordinary differential equation of \Cref{eq: GDA discrete dynamical sys}, obtained by taking infinitely small \(\gamma\), which leads to a continuous-time dynamical system
\begin{equation}\label{eq: GDA cont dynamical system}
    \dot{\z} = -\omega (\z). 
\end{equation}
Note that \(-J(\z)\) is the Jacobian of the continuous-time dynamical system in \Cref{eq: GDA cont dynamical system}. We now introduce concepts we will build upon to comment on the behavior of any algorithm used to solve \ref{eq: min x f max y f}.
\begin{definition}
    \textnormal{\textbf{(Critical point)}} Given a continuous-time dynamical system \(\dot\z = -h_c(\z)\), \(\z\in\mathbb{R}^{n+m}\) is a critical point of \(h_c\) if \(h_c(\z) = 0\). Further, if for a critical point \(\z\), \(\lambda \neq 0 \,\, \forall \,\, \lambda \in \mathrm{spec}(\nabla_\z h_c(\z))\), then \(\z\) is called a hyperbolic critical point.
\end{definition}
We can also define a similar concept for the discrete-time dynamical system counterpart.
\begin{definition}
    \textnormal{\textbf{(Fixed point)}} Given a discrete-time dynamical system \(\z_{k+1} = h_d(\z_{k}), k\geq 0\), \(\z\in\mathbb{R}^{n+m}\) is a fixed point of \(h_d\) if \(h_d(\z) = \z\).
\end{definition}
Out of the various critical and fixed point types, we are interested in locally asymptotically stable equilibria (LASE) because they are the only locally exponentially attractive hyperbolic points under the dynamics flow. This means that any dynamical system that starts close enough to a LASE point will converge to that point. 
\begin{definition} \label{definition: continuous time LASE}
    \textnormal{\textbf{(Continuous-time LASE)}} A critical point \(\z^*\in \mathbb{R}^{n+m}\) of \(h_c\)  is a LASE of the continuous-time dynamics \(\dot\z = -h_c(\z)\) if \(\mathrm{Re}(\lambda) > 0 \,\, \forall \,\, \lambda \in \mathrm{spec}(\nabla_\z h_c(\z^*))\).
\end{definition}
\begin{definition} \label{definition: discrete time LASE}
    \textnormal{\textbf{(Discrete-time LASE)}} A fixed point \(\z^*\in \mathbb{R}^{n+m}\) of \(h_d\) is a LASE of the discrete-time dynamics \(\z_{k+1} = h_d(\z_{k}), k\geq 0\) if \(\rho(\nabla_\z h_d(\z^*))< 1\), where \(\rho(A)\) denotes the spectral radius of some matrix \(A\).
\end{definition}
\subsection{Motivation: Limiting behavior of GDA}
To motivate our work, we provide an overview of key results that analyze how GDA performs when applied to \ref{eq: min x f max y f} \citep{balduzzi2018mechanics,mazumdar2019finding,mazumdar2020gradient}. If GDA converges to a hyperbolic point \(\z_{\mathrm{GDA}}\), GDA must have converged to a LASE. Thus, from definition \ref{definition: continuous time LASE},
\begin{equation}
    \mathrm{Re}(\lambda) > 0 \,\, \forall \,\, \lambda \in \mathrm{spec}\Bigg(\underbrace{\begin{bmatrix}
        \nabla^2_{\x\x}f(\x_{\mathrm{GDA}}, \y_{\mathrm{GDA}}) & \nabla^2_{\x\y}f(\x_{\mathrm{GDA}}, \y_{\mathrm{GDA}}) \\
        - \nabla^2_{\y\x}f(\x_{\mathrm{GDA}}, \y_{\mathrm{GDA}}) & - \nabla^2_{\y\y}f(\x_{\mathrm{GDA}}, \y_{\mathrm{GDA}})
    \end{bmatrix}}_{J(\z_{\mathrm{GDA}})}\Bigg).
\end{equation}
Clearly, if \(\z_{\mathrm{GDA}}\) happens to be a strict local Nash equilibrium, from \Cref{eq: notation omega and J}, we know that \(\nabla^2_{\x\x}f(\x_{\mathrm{GDA}}, \y_{\mathrm{GDA}}) \succ 0\) and \(\nabla^2_{\y\y}f(\x_{\mathrm{GDA}}, \y_{\mathrm{GDA}}) \prec 0\). Hence, from definition \ref{definition: continuous time LASE}, it is clear that \emph{all} strict local Nash equilibria of \ref{eq: min x f max y f} are LASE of the GDA dynamics. However, the converse cannot be guaranteed, and thus, a LASE point to which GDA converges may \emph{not} be a local Nash equilibrium.

Let us further examine the structure of \(J\):
\begin{equation}\label{eq: structure of J}
    J(\z) = \begin{bmatrix}
        A & B\\ -B^{\top} & D
    \end{bmatrix}, \forall \, \, \z \in \mathbb{R}^{n+m}.
\end{equation}

Only two previous works, LSS \citep{mazumdar2019finding} and CESP \citep{adolphs2019local}, leverage this structure and propose dynamical systems that have \emph{only} strict local Nash equilibria as their LASE. 
However, only the convergence rates of LSS have been analyzed, and all rate analysis is limited to the convex-concave setting.
Further, neither of these methods discusses the constrained case, which arises in many practical situations.

This motivates us to develop a novel second-order method with a dynamical system that guarantees that only strict local Nash equilibria constitute its LASE points, generalizes to the constrained settings, and has established convergence rates for the nonconvex-nonconcave settings.

%% file: sections/03_methods.tex
\renewcommand{\algorithmicrequire}{\textbf{Input:}}
\renewcommand{\algorithmicensure}{\textbf{Initialize:}}

\section{Our Method and Main Results}\label{section: 3 method}
We are now ready to show our main results. We begin with the unconstrained setting and then move to the constrained setting. All proofs not included in the main text are given in Appendix \ref{appendix: proof}.
\subsection{Unconstrained Setting}
We list some common assumptions we make for the unconstrained case below.
\begin{assumption}\label{assumption: 1 f is C3}
    The objective function \(f \in \mathcal{C}^3\).
\end{assumption}
\begin{assumption}\label{assumption: 2 hessians invertible}
    \(J(\z), \nabla^2_{\x\x}f(\x,\y),\nabla^2_{\y\y}f(\x, \y)\) are invertible at all \(\z\) where \(\omega(\z) = 0\). 
\end{assumption}
\begin{assumption}[Relaxed in \Cref{appendix: note on assumptions}]\label{assumption: 3 J top w not 0}
    \(\om\) does not belong to the null space of \(J(\z)^\top\), for all \(\z\in\mathbb{R}^{n+m}\) where $\om\neq 0$, i.e., $\om\neq 0\implies J(\z)^\top\om\neq0$.
\end{assumption}

Assumptions \ref{assumption: 1 f is C3} and \ref{assumption: 2 hessians invertible} are standard in the literature (for example, in \cite{adolphs2019local,Azizian2024,mazumdar2019finding}). Because we propose second-order methods, Assumption \ref{assumption: 1 f is C3} ensures that the objective offers meaningful first and second-order derivatives. Assumption \ref{assumption: 2 hessians invertible} ensures that the Jacobians of any dynamical system introduced in the paper can be analyzed at a critical/fixed point. \\
We make Assumption 3 for an ease in exposition of our central arguments. Similar assumptions have been made in prior work, and it has been shown that Assumption 3 can be relaxed while retaining guarantees of convergence to only a strict local Nash equilibrium \citep{mazumdar2019finding}. We will show in \Cref{appendix: note on assumptions} that Assumption \ref{assumption: 3 J top w not 0} can be relaxed for our work in a similar manner.
\par{\textbf{Motivation.}} We first introduce a continuous-time dynamical system that employs second-order derivative information, for which we can establish desirable properties and which motivates our main method. Consider the system:
\begin{equation}\label{eq: our cont time dyn sys}
    \dot{\z} = -g_c(\z) = -\left[J(\z)^{\top} J(\z) \left(J(\z)+J(\z)^{\top}\right)+E_c(\z)\right]^{-1} J(\z)^{\top}\omega(\z),
\end{equation}
where \(E_c(\z)\) is a regularization matrix chosen such that \(J(\z)^{\top} J(\z) \left(J(\z)+J(\z)^{\top}\right)+E_c(\z)\) is invertible, and \(\om = 0 \implies E_c(\z) = 0\).
Under assumptions \ref{assumption: 1 f is C3}, \ref{assumption: 2 hessians invertible}, \ref{assumption: 3 J top w not 0}, and \ref{assumption: 4 f, w lipschitz} all solutions of \Cref{eq: our cont time dyn sys} converge only to a strict local Nash equilibrium in the unconstrained setting of \ref{eq: min x f max y f}. This is because strict local Nash equilibria of \ref{eq: min x f max y f} are the \emph{only} LASE points of \Cref{eq: our cont time dyn sys}. To prove this, we first show that critical points of \(g_c\) and \(\om\) are the same.
\begin{lemma}\label{lemma: critical points of gc <=> GDA}
   Under Assumptions \ref{assumption: 1 f is C3}, \ref{assumption: 2 hessians invertible} and \ref{assumption: 3 J top w not 0}, the critical points of \(g_c\) are exactly the critical points of the GDA dynamics \(\dot\z = -\omega(\z)\).
\end{lemma}

Lemma \ref{lemma: critical points of gc <=> GDA} establishes that at every LASE \(\z\) of \Cref{eq: our cont time dyn sys}, \(\om = 0\). This helps us to prove that \Cref{eq: our cont time dyn sys} converges to only a strict local Nash equilibrium.
\begin{theorem}\label{theorem: Cont Dyn Sys LNE <=> LASE}
    Under Assumptions \ref{assumption: 1 f is C3}, \ref{assumption: 2 hessians invertible} and \ref{assumption: 3 J top w not 0}, \(\z\) is a LASE point of \(\dot \z = -g_c(\z)\) if and only if \(\z\) is a strict local Nash equilibrium of \ref{eq: min x f max y f}.
\end{theorem}
\begin{remark}
    \textnormal{\textbf{(Avoiding rotational instability)}} It is well documented that oscillations around equilibria are caused if the Jacobian of the gradient dynamics has eigenvalues with dominant imaginary parts near equilibria~\citep{mescheder2017numerics, balduzzi2018mechanics, gidel2019negative, Wang*2020On, mazumdar2019finding}. Corollary \ref{corollaly: symmetric cont case} establishes that this cannot happen for the dynamics $\dot\z = -g_c(\z)$ in \Cref{eq: our cont time dyn sys}. 
\end{remark}
\begin{corollary}\label{corollaly: symmetric cont case}
    Under Assumptions \ref{assumption: 1 f is C3} and \ref{assumption: 2 hessians invertible},  if \(\z\) is a strict local Nash equilibrium of \(g_c\), then the Jacobian \(\nabla g_c\) has only real eigenvalues at \(\z\).
\end{corollary}

\paragraph{Practical Considerations.} Although the continuous-time dynamical system we introduce in \Cref{eq: our cont time dyn sys} has desirable theoretical properties, it is not yet a practical algorithm that can solve \ref{eq: min x f max y f}. To solve \ref{eq: min x f max y f}, we require a discrete-time dynamical system. Inspired from \Cref{eq: our cont time dyn sys}, we propose \textbf{D}iscrete-time \textbf{N}ash \textbf{D}ynamics (\texttt{DND}):
\begin{equation}\label{eq: our dis dyn sys}
\begin{split}
    \z_{k+1} &= g_d(\z_k) \\ &= \z_k - \alpha_k \left(\left[J(\z_k)^{\top} J(\z_k) \left(J(\z_k)+J(\z_k)^{\top} + \beta(\z_k)\right)+E(\z_k)\right]^{-1} \right)J(\z_k)^{\top}\omega(\z_k).  
\end{split}
\end{equation}
Regularization \(E(\z_k)\) is chosen to maintain invertibility in \Cref{eq: our dis dyn sys} and adheres to the condition that \(\omega(\z_k) = 0 \implies E(\z_k) = 0\). In contrast to the continuous-time system \(g_c\) in \Cref{eq: our cont time dyn sys}, \texttt{DND} in \Cref{eq: our dis dyn sys} contains an extra regularization term \(\beta(\z_k)\). Adding \(\beta(\z_k)\) guarantees the stability of \Cref{eq: our dis dyn sys} in accordance with Definition \ref{definition: discrete time LASE}, and is given by
\begin{equation}\label{eq: choosing beta for disc dyn sys}
   \beta(\z) = \begin{bmatrix}
        \mathbbm{1}_{\{\lambda_\x > 0\}}(b_\x)I & 0 \\ 0 & \mathbbm{1}_{\{\lambda_\y < 0\}}(b_\y)I
    \end{bmatrix},
\end{equation}
where \(\lambda_\x\) and \(\lambda_\y\) denote the minimum and maximum eigenvalues of \(\nabla^2_{\x\x} f(\x, \y)\) and \(\nabla^2_{\y\y} f(\x, \y)\) respectively. These eigenvalues can be found through computations involving Hessian-vector products, which can be made as efficient as gradient evaluations \citep{pearlmutter1994fast, lanczos1950iteration}. The terms \(b_\x\) and \(b_\y\) can be taken to be any constants as long as \(b_\x > \nicefrac{1}{2}\) and \(b_\y < \nicefrac{-1}{2}\).

\(\beta(\z)\) is a non-smooth regularization term, but it is differentiable around any fixed point of \(\omega\). The following theorem shows that \texttt{DND} inherits all the desirable properties that we established for the continuous-time system \(g_c\).

\begin{theorem}\label{theorem: Discrete time dyn sys <=> LNE}
    Under Assumptions \ref{assumption: 1 f is C3}, \ref{assumption: 2 hessians invertible}, and \ref{assumption: 3 J top w not 0}, for any \(\alpha_k\in(0,1]\), \texttt{DND}, with \(\beta(\z)\) chosen as in \Cref{eq: choosing beta for disc dyn sys} satisfies the following:
    \begin{enumerate}
        \item The fixed points of \texttt{DND} are exactly the fixed points of the discrete-time GDA dynamics in \Cref{eq: GDA discrete dynamical sys}.
        \item \(\z\) is a LASE of \texttt{DND} \(\iff\) \(\z\) is a strict local Nash equilibrium of unconstrained \ref{eq: min x f max y f}.
        \item If \(\z\) is a fixed point of \texttt{DND}, then the Jacobian \(\nabla g_d\) has only real eigenvalues at \(\z\).
    \end{enumerate}
\end{theorem}
We will prove in \Cref{section: convergence rates} that if \texttt{DND} converges, it converges to a local Nash equilibrium with a \emph{linear} local asymptotic convergence rate. In the remainder of this section, we consider a modification to \texttt{DND} which can yield useful speed-ups in convergence in certain settings.

\paragraph{Can we speed up \texttt{DND}?} We motivate a modification to \Cref{eq: our dis dyn sys}, which allows for superlinear convergence to a ball-shaped region around a fixed point. If this fixed point is a LASE of \texttt{DND} (and therefore also a strict local Nash equilibrium of \ref{eq: min x f max y f} by \Cref{theorem: Discrete time dyn sys <=> LNE}), the modification can achieve rapid convergence to a small neighborhood of this local Nash point. 
The modification retains desirable stability guarantees and escapes the ball if the fixed point is not a LASE. 
The radius of the ball can be treated as a hyperparameter and tuned for good performance. 

\paragraph{Modified discrete-time system.} We call the modified method \textbf{Sec}ond \textbf{O}rder \textbf{N}ash \textbf{D}ynamics (\second), which is given by
\begin{equation}\label{eq: our modified dis dyn sys}
    \z_{k+1} = \begin{cases}
        z_k - \alpha_k \left(S(\z_k)\right)^{-1}J(\z_k)^{\top}\omega(\z_k),&\lVert \z_k - \z_{k-1}\rVert > \epsilon \\ 
        g_d(\z_k),& \mathrm{ else.}
    \end{cases}
\end{equation}
where \(\epsilon >0\) is a user-specified constant, the matrix \(S(\z_k) \succ 0\) and can be derived from modifying the term \(J(\z_k)^{\top} J(\z_k) \left(J(\z_k)+J(\z_k)^{\top} + \beta(\z)\right)\) with an appropriate regularization \(E(\z_k)\) in \Cref{eq: our dis dyn sys}. We define such a choice in Appendix \ref{appendix: choice of S in SecOND}. 

\paragraph{Reinterpretation as a Gauss-Newton method far from fixed points.} 
Consider the problem
\begin{equation}\label{eq: least sq interpret}
    \min_{\z\in\mathbb{R}^{n+m}} \underbrace{\frac{1}{2}\|\om\|^2_2}_{\ell(\z)}.
\end{equation}
We observe that \(\nabla_\z \ell(\z) = J(\z)^\top \om\). For \(\|\z_k - \z_{k-1}\| > \epsilon\), we have the system \(\z_{k+1} = \z_k - (S(\z_k))^{-1}\nabla_\z \ell(\z),\) with \(S(\z_k) \succ 0\), which is a \textit{modified Gauss-Newton} algorithm for solving \Cref{eq: least sq interpret}. By choosing \(S(\z_k)\approx J(\z_k)^\top J(\z_k)\) (see Appendix \ref{appendix: choice of S in SecOND}), if the Gauss-Newton system converges to a fixed point \(\z_c\), we can be guaranteed a \emph{local superlinear rate of convergence to that point}. 
Moreover, whenever \(\|\z_k - \z_{k-1}\| > \epsilon\), we may choose step size \(\alpha_k\) according to any standard line search rule from nonlinear programming  \citep{nocedal1999numerical, bertsekas1997nonlinear}.
For example, in our implementation, we choose a backtracking line search with the Armijo condition \citep{armijo1966minimization} and choose an \(\alpha_k\) for some \(c\in(0,1)\) such that
\begin{equation}\label{eq: armijo}
    \ell(\z_k) - \ell(\z_{k+1}) \geq c \alpha_k \omega(\z_{k})^\top J(\z_{k}) (S(\z_k))^{-1}J(\z_{k})^\top \omega(\z_{k}).
\end{equation}
 Based on \second, we construct Algorithm \ref{algo: modified discrete}, which converges with a local quadratic rate toward the first fixed point it encounters, and switches to \texttt{DND} when it is close enough to that point. 
 If the fixed point satisfies the strict local Nash equilibrium sufficiency conditions given in Definition \ref{definition: DNE}, \texttt{SecOND} will have reached the point faster than \texttt{DND} would have from the same initialization. If the fixed point does not satisfy strict local Nash conditions, the switch to \texttt{DND} dynamics ensures that the iterates avoid convergence to the spurious fixed point. 
 More sophisticated variants which allow for switching back and forth multiple times can also be considered. We will establish that  \texttt{SecOND} inherits the desirable stability properties of \texttt{DND}, and approaches a fixed point at a comparatively faster, quadratic local rate in \Cref{section: convergence rates}.
\begin{algorithm}
\caption{\textbf{Sec}ond \textbf{O}rder \textbf{N}ash \textbf{D}ynamics (\texttt{SecOND})}\label{algo: modified discrete}
\begin{algorithmic}
\Require Functions \(\omega(\z), J(\z), S(\z)\); initial point \(\hat\z\); constants \(\epsilon>0, 0<\alpha_0\leq1\)
\Ensure \(\z_0 \gets \hat\z, \z_{1} \gets \z_0 - \alpha_0 (S(\z_0))^{-1}J(\z_0)^{\top}\omega(\z_0),  k = 1\)
\While{not converged}
\If{\(\lVert \z_{k} -\z_{k-1}\rVert > \epsilon\)}
    \State Choose \(\alpha_k\) with appropriate line search \Comment{for example, from \Cref{eq: armijo}}
    \State \(\z_{k+1} \gets z_k - \alpha_k (S(\z_k))^{-1}J(\z_k)^{\top}\omega(\z_k)\)  \Comment{from \Cref{eq: our modified dis dyn sys}}
\ElsIf{\(\z_k\) does not satisfy strict LNE sufficiency conditions} \Comment{from Definition (\ref{definition: DNE})}
    \State \(\z_{k+1} \gets g_d(\z_k)\) \Comment{from \Cref{eq: our dis dyn sys}}
\Else{} \State \textbf{break}
\EndIf
\State \(k\gets k + 1\)
\EndWhile \\ 
\Return \(\z_k\)
\end{algorithmic}
\end{algorithm}

\subsection{Constrained Setting}
\paragraph{Notation.} \(\Pi_{\mathcal{Q}}[\mathbf{p}]\) denotes the Euclidean projection of some vector \(\mathbf{p}\) onto some set \(\mathcal{Q}\). \(\mathrm{proj}_{\mathbf{a}}(\mathbf{b})\) denotes the Euclidean projection of a vector \(\mathbf{b}\) onto another vector \(\mathbf{a}\). $\intg$ and $\boundg$ denote the interior and boundary of \(\gset\) respectively.

Intuitively, any local generalized Nash equilibrium in $\intg$ is actually also a strict local Nash equilibrium of the unconstrained game. Therefore, if the Euclidean projections of the \texttt{DND} iterates converge to a point in \(\mathrm{int\,}\mathcal{G}\), this point must be a local generalized Nash equilibrium. Further, if a step taken by \texttt{DND} at a point \(\z\) in $\boundg$ is parallel to \(-\om\), then, from Definition \ref{definition: GNE}, \(\z\) is a local Generalized Nash equilibrium as well.
\paragraph{Algorithm for Constrained Setting.} Based on the above discussion, we construct \textbf{Se}cond-order \textbf{Co}nstrained \textbf{N}ash \textbf{D}ynamics (\texttt{SeCoND}), given in Algorithm \ref{algo: constrained}, for solving a constrained \ref{eq: min x f max y f}. \texttt{SeCoND} has the property that if it converges, it converges to a local Generalized Nash equilibrium that follows Definition \ref{definition: GNE}. 
If desired, the convergence of Algorithm \ref{algo: constrained} can be accelerated via a Gauss-Newton approach analogous to \Cref{eq: our modified dis dyn sys}.

\begin{assumption}\label{assumption: 5 convex constraints}
    The set \(\mathcal{G}\) is convex.
\end{assumption}
Assumption \ref{assumption: 5 convex constraints} has been shown to hold for several problems of practical interest \citep{facchinei2010generalized}.
\begin{theorem}\label{theorem: constrained case}
    Let Assumptions \ref{assumption: 1 f is C3}, \ref{assumption: 2 hessians invertible}, \ref{assumption: 3 J top w not 0}, and \ref{assumption: 5 convex constraints} hold, and let $\omega(\mathbf{z})\neq 0 ~\forall~\mathbf{z} \in\partial G$. Let \texttt{SeCoND} be initialized from a point that is not a non-Nash fixed point. Then, if SeCoND (Algorithm \ref{algo: constrained}) converges to a point $\mathbf{z}$:
    \begin{enumerate}
        \item If $\mathbf{z}\in \mathrm{int} G$, then $\mathbf{z}$ is a strict local generalized Nash equilibrium.
        \item If $\mathbf{z}\in \partial G$, then $\mathbf{z}$ is a local generalized Nash equilibrium (not necessarily strict).
    \end{enumerate}
\end{theorem}
\begin{algorithm}
\caption{\textbf{Se}cond-order \textbf{Co}nstrained \textbf{N}ash \textbf{D}ynamics (\texttt{SeCoND})}\label{algo: constrained}
\begin{algorithmic}
\Require Functions \(\omega(\z), J(\z);\) set \(\mathcal{G};\) initial point \(\hat{\z}\); constant \(\alpha\) 
\Ensure \(\z_0 \gets \hat{\z}, k = 0\)
\While{not converged}
\If{\(\z_k \in \mathrm{Int\,}\mathcal{G}\)}
    \State \(\z_{k+1} \gets \Pi_{\mathcal{G}}\left[g_d(\z_k)\right]\)  \Comment{from \Cref{eq: our dis dyn sys}}
\ElsIf{\(\z_k \in \partial\mathcal{G}\)}  \Comment{\(E\) from \Cref{eq: our dis dyn sys}, \(\beta\) from \Cref{eq: choosing beta for disc dyn sys}}
    \State \(\mathbf{m} \gets \mathrm{proj}_{\omega(\z_k)}\left(\left[J(\z_k)^{\top} J(\z_k) \left(J(\z_k)+J(\z_k)^{\top} + \beta(\z_k)\right)+E(\z_k)\right]^{-1} J(\z_k)^{\top}\omega(\z_k)\right)\)
    \State \(\z_{k+1} \gets \Pi_{\mathcal{G}}\left[\z_k - \alpha \mathbf{m}\right]\) 
\EndIf
\State \(k\gets k + 1\)
\EndWhile \\ 
\Return \(\z_k\)
\end{algorithmic}
\end{algorithm}

%% file: sections/04_Rates.tex
\section{Rates of Convergence in Unconstrained Games with Different Structures}\label{section: convergence rates}
We investigate the convergence rate of $\texttt{SecOND}$ in unconstrained zero-sum games satisfying a variety of structural assumptions. We find that the modification introduced in \Cref{eq: our modified dis dyn sys} enables \texttt{SecOND} to have competitive last-iterate rates in bilinear and convex-concave games, where converging to a LASE point of the GDA dynamics guarantees convergence to local Nash equilibria. For nonconvex-nonconcave games, we analyze the \texttt{DND} algorithm---our core contribution---and we find that if \texttt{DND} converges to a local Nash equilibrium, it does so with a linear local asymptotic convergence rate. We then show that the unconstrained \texttt{SecOND} algorithm maintains the desirable properties of \texttt{DND}, while also approaching a fixed point with a local quadratic rate.

\subsection{Bilinear Zero-Sum Games}
We consider the class of bilinear zero-sum games, which are a form of convex-concave games with the objective $f(\x,\y)$ in \ref{eq: min x f max y f} taking the form
\begin{align}
    f(\x, \y) = \x^\top A \y. \label{eq: bilinear}
\end{align}
It is well known that any LASE point of the GDA dynamics, when employed for \Cref{eq: bilinear} must be a global Nash equilibrium. Thus, one can use \Cref{algo: modified discrete} using only the Gauss-Newton iterates given in \Cref{eq: our modified dis dyn sys} to find a Nash equilibrium.

\begin{theorem}\label{theorem: bilinear}
    Assume that $J(\z)$ is invertible, then for step size $0<\alpha<1$, \texttt{SecOND} yields a problem independent global linear last literate convergence rate for the bilinear zero sum game given in \Cref{eq: bilinear}, with iterates $\z_k,k\geq 0$ satisfying
    \begin{align*}
        \|\z_k\| \leq (1-\alpha)^k \|\z_0\|.
    \end{align*}
\end{theorem}

\begin{proof}
    For games of the form given in \Cref{eq: bilinear}, \texttt{SecOND} dynamics become
    \begin{align*}
        \z_{k+1} &= \z_k - \alpha_k\left[J(\z_k)^\top J(\z_k)\right]^{-1}J(\zk)^\top \omega(\z)\\
        &=\z_k - \alpha_k\left[J(\z_k)^\top J(\z_k)\right]^{-1}J(\zk)^\top J(\zk)\zk \tag{for $f=\x^\top A\y, J(\z)=\omega(\z)\z$}\\
        &=(1-\alpha_k)\zk
    \end{align*}
    Choosing $\alpha_k = \alpha < 1 ~\forall~k\geq 0$, we get a linear rate of convergence, which is independent of the problem objective $f$, and depends solely on the step size $\alpha$, as
    \begin{align*}
        \|\z_k\| \leq (1-\alpha)^k \|\z_0\|.
    \end{align*}
\end{proof}
The assumption of invertibility of $J(\z)$ in \Cref{theorem: bilinear} is equivalent to the assumption that $A$ in \Cref{eq: bilinear} is a square, full rank matrix. We note that under this assumption, prior works have also established linear last-iterate convergence rate for many popular first-order methods -- proximal point, optimistic gradient descent-ascent, and extragradient methods \citep{rockafellar1976monotone, liang2019interaction, mokhtari2020unified}. However, we reiterate that in less structured settings beyond bilinear games, all these first order methods suffer from issues like cycling, or convergence to non-local Nash equilibria. The closely related work LSS, which is also a second order method and enjoys continuous-time guarantees of convergence to only strict local Nash equilibria, also reports a linear last-iterate convergence rate for bilinear games with full rank, square matrices \citep{mazumdar2019finding}. 
\subsection{Convex-Concave Zero-Sum Games}
In this section, we consider the last-iterate convergence of \texttt{SecOND} in convex-concave games that are more general than bilinear games, and have second-order information that can be leveraged by \texttt{SecOND}. To this end, we make the following assumption:
\begin{assumption}\label{assumption: sufficient bilinear}
    The convex-concave games we consider satisfy $J(\z)^\top J(\z) \succeq \mu^2 I$ for some $\mu > 0$.
\end{assumption}
Assumption \ref{assumption: sufficient bilinear} is common in works that investigate second order algorithms in zero-sum games, with various versions appearing in literature \citep{azizian2020tight, abernethy2021last, lu2022sr, grimmer2023landscape, mazumdar2019finding}.
In particular, Assumption \ref{assumption: sufficient bilinear} can be thought of as a relaxation on claiming that the zero-sum game in consideration is strongly monotone. A strongly monotone zero-sum game has a strongly convex-strongly concave objective, and has a unique global Nash equilibrium. In comparison, Assumption \ref{assumption: sufficient bilinear} allows us to admit convex-concave games for which multiple isolated Nash equilibria might occur. 

For convex-concave games, finding the fixed points of the GDA dynamics is sufficient to guarantee finding a local Nash equilibrium, thus in this section, we again analyse the performance of \texttt{SecOND} utilizing only Gauss-Newton iterates from \Cref{eq: our modified dis dyn sys}. We also make assumptions on the smoothness of $\ell(\z)$, lipschitzness of $\om$ and $J(\z)$, which are standard in literature (for example, in \cite{adolphs2019local,Azizian2024,mazumdar2019finding}.
\begin{assumption} \label{assumption: ell square smooth}
    The function $\ell(\z) = \frac{1}{2}\|\om\|^2$ is $L-$smooth.
\end{assumption}
\begin{assumption}\label{assumption: 4 f, w lipschitz}
    \(\omega\) is \(L_\omega\)-Lipschitz, and \(J\) is \(L_J\)-Lipschitz.
\end{assumption}
We find that for convex-concave games, \texttt{SecOND} has global linear and local superlinear last-iterate convergence rates.

\begin{theorem}\label{theorem: convex concave conv rate}
    Assume that \ref{eq: min x f max y f} is convex-concave. Then, we have that
    \begin{enumerate}
        \item Let Assumptions \ref{assumption: 1 f is C3}, \ref{assumption: 3 J top w not 0}, \ref{assumption: sufficient bilinear} and \ref{assumption: ell square smooth} hold, then, for a constant step size $0<\alpha <\min\left\{\nicefrac{1}{2}, \nicefrac{\mu^2}{L}\right\}$, \texttt{SecOND} exhibits a global linear last-iterate convergence rate with
        \begin{align*}
            \|\omega(\zkone)\|^2 &\leq \underbrace{\left(1 - 2\alpha_k + \alpha_k^2 \frac{L}{\mu^2} \right)}_{<\,1~\textrm{for}~0<\alpha <\min\left\{\frac{1}{2}, \frac{\mu^2}{L}\right\}}\|\omega(\zk)\|^2, ~\forall~k=0,1,2,\dots.
        \end{align*}
        \item Under Assumptions \ref{assumption: 1 f is C3}, \ref{assumption: 2 hessians invertible}, \ref{assumption: 3 J top w not 0}, \ref{assumption: sufficient bilinear}, and \ref{assumption: 4 f, w lipschitz}, let $\z^*$ be the local Nash equilibrium to which \texttt{SecOND} converges, when initialized at $\z_0$. Then, \texttt{SecOND} exhibits a local quadratic last-iterate convergence rate with
        \begin{align*}
           \|\zkone-\z^*\| \leq \frac{L_J}{2\mu}  \|\zk-\z^*\|^2,~\forall~k=0,1,2,\dots.
        \end{align*}
    \end{enumerate}
\end{theorem}
\begin{proof}
    Note that Assumption \ref{assumption: sufficient bilinear} implies that $J(\z)$ is invertible for all $\z\in\mathbb{R}^{n+m}$. To show the global linear rate, we have from the smoothness of $\ell(\z)$
    \begin{align*}
       \ellfunc{\zkone} &\leq \ellfunc{\zk} - \alpha_k\omega(\zk)^\top J(\zk)\left[ J(\zk)^\top J(\zk)\right]^{-1} J(\zk)^\top \omega(\zk)\\
       &+ \frac{\alpha_k^2 L}{2} \omega(\zk)^\top J(\zk)\left[ J(\zk)^\top J(\zk)\right]^{-1}\left[ J(\zk)^\top J(\zk)\right]^{-1} J(\zk)^\top \omega(\zk)\\
       &= \left(\frac{1}{2} - \alpha_k \right) \|\omega(\zk)\|^2 + \frac{\alpha_k^2 L}{2} \omega(\zk)^\top\left[J(\zk) J(\zk)^\top \right]^{-1} \omega(\zk) \tag{invertibility of $J(\z)$}\\
       \implies \|\omega(\zkone)\|^2 &\leq \underbrace{\left(1 - 2\alpha_k + \alpha_k^2 \frac{L}{\mu^2} \right)}_{:=h(\alpha_k)}\|\omega(\zk)\|^2 \tag{$\left(J(\zk) J(\zk)^\top \right)^{-1}\preceq \frac{1}{\mu^2}I$}
    \end{align*}
    We will now show that it is always possible to choose $\alpha_k>0$ such that $0<h(\alpha) < 1$.
    Let us analyse $h(\alpha)$ by considering two cases:
    \begin{itemize}
        \item When $\nicefrac{\mu^2}{L} < 1:$ We have that $h(\alpha) > 0 ~\forall ~\alpha$, and $h$ takes a minimum value at $\alpha=\nicefrac{\mu^2}{L}$, yielding
        \begin{align*}
            h\left(\frac{\mu^2}{L}\right) = 1 - \frac{\mu^2}{L} < 1.
        \end{align*}
         Note that $h(0) = 1$, thus $h(\alpha)<1~\forall~\alpha\in\left(0,\nicefrac{\mu^2}{L}\right]$.
         \item When $\nicefrac{\mu^2}{L} \geq 1:$ In this case $h(1) = \nicefrac{L}{\mu^2}-1 \leq 0$, thus $h$ has a root in between $0$ and $1$. Note that $h(\nicefrac{1}{2}) = \nicefrac{L}{4\mu^2}<1.$ Thus $h(\alpha)<1~\forall~\alpha\in(0,0.5]$ for this case.
    \end{itemize}
    For showing local quadratic convergence we note that Assumption \ref{assumption: sufficient bilinear} implies that the spectral norm $\|J(\z)^{-1}\|\leq \nicefrac{1}{\mu}$. We have 
    \begin{align*}
        \left\|\zkone - \z^*\right\| &= \left\|\zk - \z^* - \left[ J(\zk)^\top J(\zk) \right]^{-1}J(\zk)^\top \omega(\zk)\right\|\\
        &= \left\|\left[ J(\zk)^\top J(\zk) \right]^{-1} \left( J(\zk)^\top J(\zk)\left(\zk-\z^*\right) -  J(\zk)^\top \omega(\zk)\right)\right\| \\
        &= \Bigg\|\left[ J(\zk)^\top J(\zk) \right]^{-1} \tag{from Taylor's Theorem}\\
        &\left( J(\zk)^\top J(\zk)\left(\zk-\z^*\right) -  J(\zk)^\top \int^1_0 J\left(\z^* + t(\zk - \z^*)\right)\left(\zk-\z^*\right)dt \right)\Bigg\|  \\
        &= \left\| \left[ J(\zk)^\top J(\zk) \right]^{-1} \int^1_0 \left(J(\zk)^\top J(\zk) - J(\zk)^\top J\left(\z^* + t(\zk - \z^*)\right)\right)\zk-\z^* dt \right\| \\
        &= \left\| J(\zk)^{-1} \int^1_0 \left(J(\zk) - J\left(\z^* + t(\zk-\z^*)\right)\right)\zk-\z^* dt\right\|\\
         &\leq \left\|J(\zk)^{-1}\right\| \int^1_0 \left\|J(\zk) - J\left(\z^* + t(\zk-\z^*)\right)\right\|\|\zk-\z^*\|dt\\
        &\leq L_J \left\|J(\zk)^{-1}\right\| \int^1_0 \|\zk-\z^*\|^2 t dt
    \end{align*}
\begin{align*}
    \implies \left\|\zkone - \z^*\right\| & \leq \frac{L_J}{2} \left\|J(\zk)^{-1}\right\| \|\zk-\z^*\|^2 \leq \frac{L_J}{2\mu} \|\zk-\z^*\|^2 .
\end{align*}
\end{proof}

\subsection{Nonconvex-Nonconcave Zero-Sum Games}
We now consider the most general setting of nonconvex-nonconcave zero-sum games. In this setting, prior works that establish guarantees of convergence to only strict local Nash equilibria \citep{adolphs2019local,mazumdar2019finding} \textit{do not} have established rates of convergence. This section establishes the first rates in this setting: specifically, we show that if it converges, $\texttt{DND}$ has a linear local asymptotic rate of convergence.
\begin{theorem}\label{theorem: linear convergence rate}
Assume that a strict local Nash equilibrium of \ref{eq: min x f max y f} exists. Let \texttt{DND} be initialized from a random point that is not a non-Nash fixed point, and chosen from a non-degenerate distribution. Under Assumptions \ref{assumption: 1 f is C3}, \ref{assumption: 2 hessians invertible}, \ref{assumption: 3 J top w not 0}, and \ref{assumption: 4 f, w lipschitz}, if $\texttt{DND}$ converges, it converges almost surely to a strict local Nash equilibrium of \ref{eq: min x f max y f}. Further, if the step size is chosen as $\alpha_k\leq\max\{2|\lambda_{\mathbf{x}}|, 2|\lambda_{\mathbf{y}}|\}$ then $\texttt{DND}$ has a linear local asymptotic convergence rate given by \begin{equation*}
\lim_{k\rightarrow\infty} \frac{\lVert \mathbf{z}_{k+1} - \mathbf{z}^*\rVert}{\lVert\mathbf{z}_k - \mathbf{z}^*\rVert} \leq \max \left\{ \left(1 - \frac{\alpha}{2\tilde{\lambda}_{\mathbf{x}}}\right), \left(1 + \frac{\alpha}{2\tilde{\lambda}_{\mathbf{y}}}\right)  \right\}. 
\end{equation*}

Here, $\alpha$ is the step size at the sequence limit in \Cref{eq: our dis dyn sys}, and $\lambda_{\mathbf{x}}, \lambda_{\mathbf{y}}$ refer to the quantities in \Cref{eq: choosing beta for disc dyn sys}, and $\tilde{\lambda}_{\mathbf{x}}>0, \tilde{\lambda}_{\mathbf{x}}<0$ denote $\lambda_{\mathbf{x}}, \lambda_{\mathbf{y}}$ evaluated at the sequence limit.
\end{theorem}
We now show that \texttt{SecOND} retains the desirable convergence properties of \texttt{DND}, with an accelerate rate of convergence to the first critical point it encounters. Note that this rate does not require Assumption \ref{assumption: sufficient bilinear} as in \Cref{theorem: convex concave conv rate}.
\begin{theorem}\label{theorem: modified stable}
Consider \ref{eq: min x f max y f}, when $f$ is nonconvex-nonconcave. Let $\z_c$ be a critical point of $f$. Then under Assumptions \ref{assumption: 1 f is C3}, \ref{assumption: 2 hessians invertible}, \ref{assumption: 3 J top w not 0}, and \ref{assumption: 4 f, w lipschitz}:
\begin{enumerate}
    \item $\mathbf{z}$ is a LASE of \texttt{SecOND} (Algorithm \ref{algo: modified discrete}) if and only if  $\mathbf{z}$ is a strict local Nash equilibrium of \ref{eq: min x f max y f}.
    \item Assume that a strict local Nash equilibrium of \ref{eq: min x f max y f} exists. Let \texttt{SecOND} be initialized at a random point \(\z_0\) (chosen from a non-degenerate distribution) that is not a non-Nash fixed point. If \texttt{SecOND} converges, it converges almost surely to a strict local Nash equilibrium of \ref{eq: min x f max y f}.
    \item Let $\mathcal{B}_\delta(\z_c)$ denote the ball $\{\z~\vert~\|\z-\z_c\|\leq \delta\}$. If $\z_0\in\mathcal{B}_\delta(\z_c)$, and $S(\z_k)$ is chosen as constructed in Appendix \ref{appendix: choice of S in SecOND}, then Algorithm \ref{algo: modified discrete} approaches $\mathbf{z}_c$ quadratically with a rate \begin{equation*}
    \lVert \mathbf{z}_{k+1} - \mathbf{z}_c\rVert \leq \frac{ML_\omega L_J}{2} \lVert \mathbf{z}_{k} - \mathbf{z}_c\rVert^2, \forall~k = 0, 1, \dots,\text{ where $M = \sup_{z\in \mathcal{B}_\delta(\z_c)} \lVert S(\z)^{-1} \rVert$},
\end{equation*}
until the dynamics are switched to \texttt{DND}.
\end{enumerate} 

\end{theorem}
In the \texttt{SecOND} algorithm, once the dynamics are switched to \texttt{DND}, \Cref{theorem: linear convergence rate} applies and ensures that either (i) the iterates converge to $\z_c$ if it is a strict local Nash equilibrium (with a asymptotically local linear rate), or(ii) escape the neighborhood of $\z_c$ if it is a non local Nash equilibrium.

%% file: sections/05_exps.tex
\section{Experiments}\label{section: exp}
We now investigate how well the theoretical properties of our algorithms transfer to practical problems. Our main aims are: (i) to compare the performance of \texttt{SecOND} with previous related work in unconstrained, nonconvex-nonconcave settings, (ii) to determine if modifications made to \texttt{DND} in \texttt{SecOND} are beneficial, (iii) to test whether \texttt{SeCoND} converges to a local generalized Nash equilibrium in the constrained setting, (iv) to test the performance of \texttt{SeCoND} in constrained problems with empty interiors. All details of the experimental setup are included in Appendix \ref{Appendix: Exp details}.
\subsection{Two-Dimensional Toy Example}
We consider the function 
\begin{equation*}
    f(x, y) = e^{-0.01(x^2+y^2)}((0.3x^2+y)^2+(0.5y^2+x)^2), \, x, y\in \mathbb{R}.
\end{equation*}
This function is nonconvex-nonconcave, and the unconstrained version of \ref{eq: min x f max y f} has three local Nash equilibria, while the GDA dynamical system in \Cref{eq: GDA cont dynamical system} has 4 LASE points for this function. 
\paragraph{Baselines.} In this experiment, we tested the performance of \texttt{SecOND} (Algorithm \ref{algo: modified discrete}) against three baselines: \texttt{DND}, gradient descent-ascent (GDA), and local symplectic surgery (LSS) \citep{mazumdar2019finding}, on 10000 random initializations. 

\subsubsection{Does \texttt{SecOND} provide faster convergence than baselines?}
Figure \ref{fig: Unconstrained Toy}(a) shows the difference in the number of iterations taken to converge within a fixed tolerance by \texttt{SecOND} and each respective baseline. \texttt{SecOND} consistently converged more rapidly than LSS, achieving a still greater performance improvement than GDA.
Finally, we note that we could not compare to the CESP method \citep{adolphs2019local}, because it could not reliably converge in our experiments (see Appendix \ref{appendix: toy baselines}). An additional experiment investigating convergence of all algorithms to a local Nash equilibrium is in Appendix \ref{appendix: additional result}.
\subsubsection{Does \texttt{SecOND} perform better than \texttt{DND}?} 
\begin{figure}[t]
\begin{subfigure}{\textwidth}
  \centering
  \includegraphics[width=8cm]{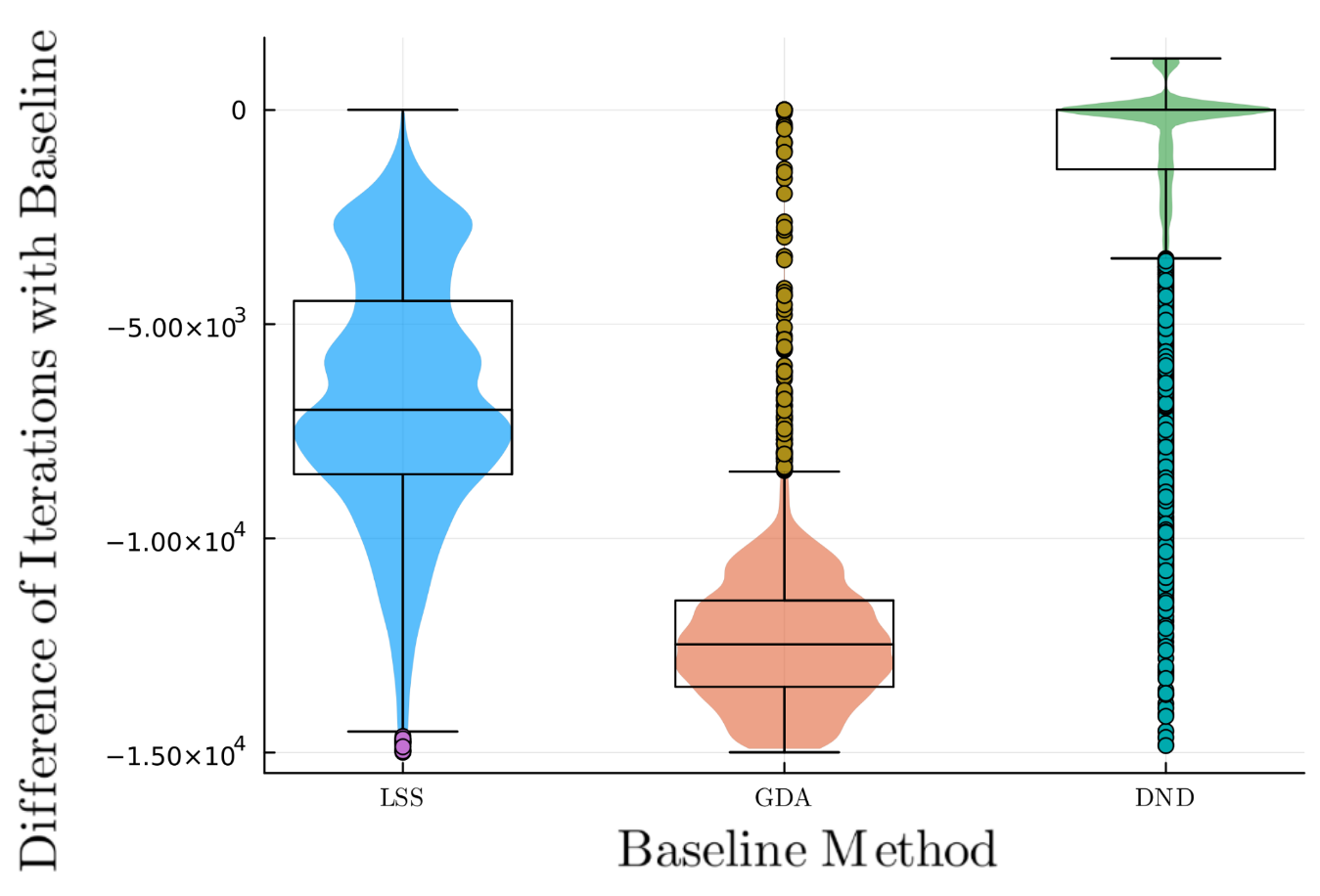}
    \caption{Violin plot of the difference in iterations taken between \texttt{SecOND} and each baseline method (lower is better).  \texttt{SecOND} converges faster than baselines for the unconstrained \ref{eq: min x f max y f}. Dots represent outliers, (see Appendix \ref{appendix: toy ex params}).}
\end{subfigure}
\begin{subfigure}{.5\textwidth}
  \centering
  \includegraphics[width=0.8\linewidth]{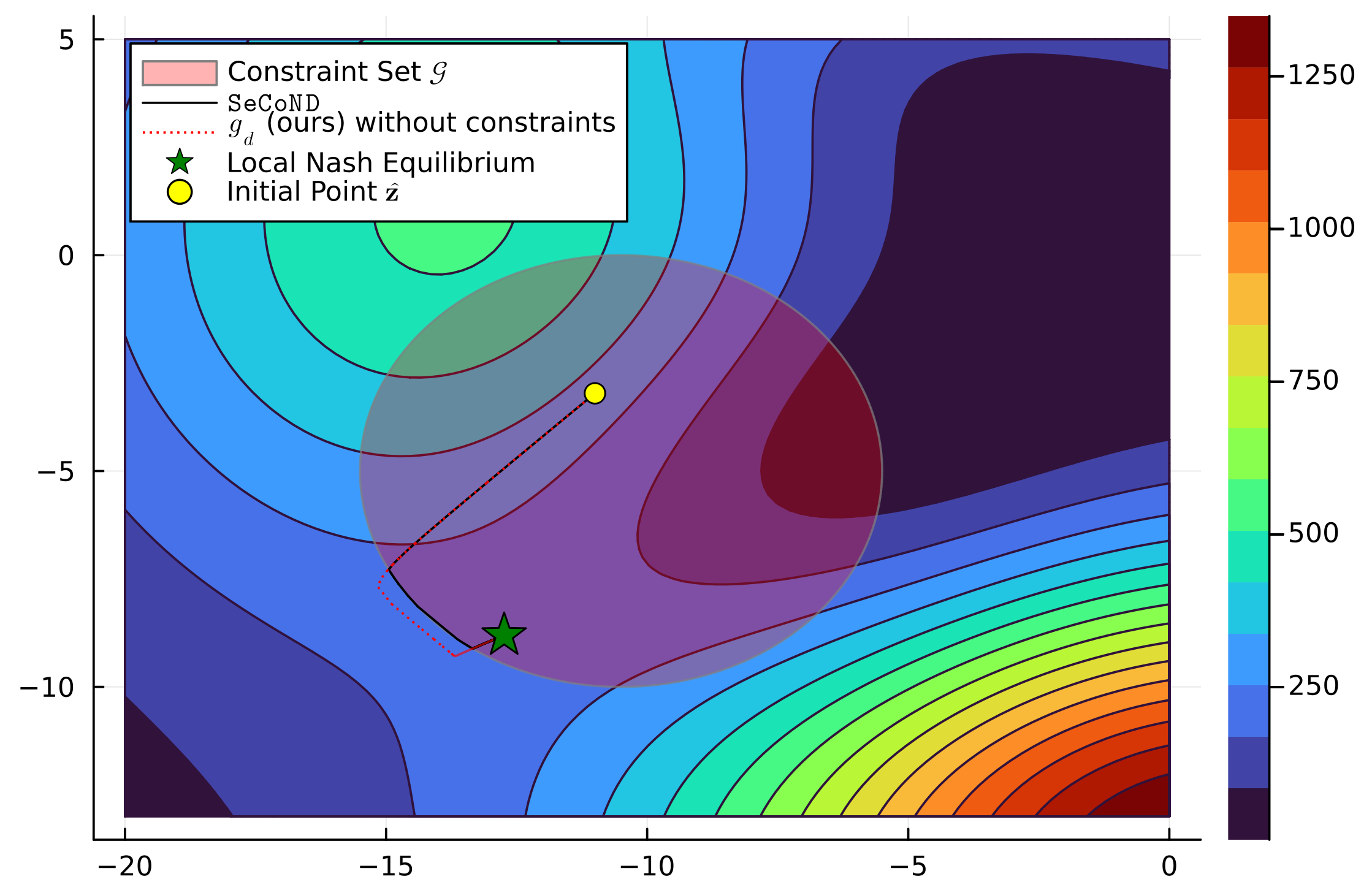}
  \caption{\texttt{SeCoND} in Constrained \ref{eq: min x f max y f}.}
\end{subfigure}
\begin{subfigure}{.5\textwidth}
 \centering
  \includegraphics[width=0.8\linewidth]{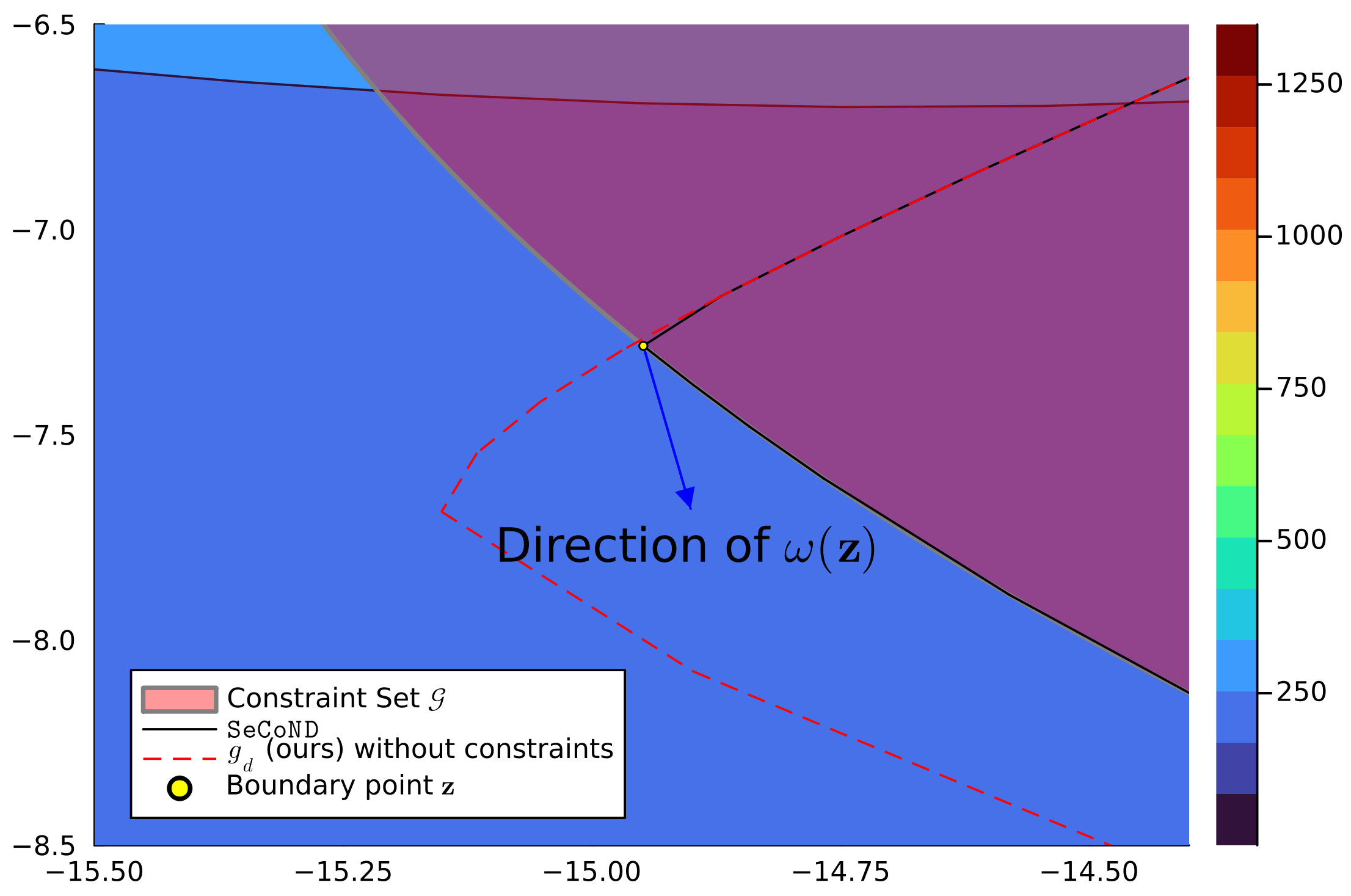}
  \caption{Geometry where \texttt{SeCoND} first hits the boundary.}
\end{subfigure}
\caption{Numerical results for a two-dimensional toy example.}
\label{fig: Unconstrained Toy}
\end{figure}

From Figure \ref{fig: Unconstrained Toy}(a), we observe that \texttt{SecOND} performed similarly to \texttt{DND} in this numerical example. \texttt{DND} outperformed \texttt{SecOND} in some instances, which occurred when \texttt{SecOND} initially went to the neighborhood of an undesirable critical point, at which %
the quantity \(\|\om\|^2_2 \approx 0\). %
In such cases, \texttt{SecOND} had to correct its course to go to the desirable fixed points. This made it converge slower than \texttt{DND}, which went to the desirable fixed points in the first place. In the cases when \texttt{SecOND} rapidly approaches a desirable critical point, \texttt{SecOND} converged much faster than \texttt{DND}. This shows that the modification made to \texttt{DND} in \texttt{SecOND} can indeed be advantageous. 
\subsubsection{Does \texttt{SeCoND} converge to a local generalized Nash equilibrium?} We tested \texttt{SeCoND} (Algorithm \ref{algo: constrained}) in this toy setting by including a constraint of the form \((x+10.5)^2 + (y+5)^2 \leq 25\),
and found that \texttt{SeCoND} successfully converges to a local generalized Nash equilibrium. As seen in Figure \ref{fig: Unconstrained Toy}(b), \texttt{SeCoND} initially follows \texttt{DND} while iterates remain in the interior of the feasible set. However, after hitting the boundary, \texttt{SeCoND} remains on the boundary before returning to the interior and converging to the same local (generalized) Nash equilibrium as \texttt{DND}. Figure \ref{fig: Unconstrained Toy}(c) is representative of the geometry across the portion where \texttt{SeCoND} remains on the boundary. Because \(-\om\) is not parallel to the constraint gradient here, \texttt{SeCoND} eventually returns to the interior. 
\begin{figure}[t]
    \centering
    \includegraphics[width=0.9\linewidth]{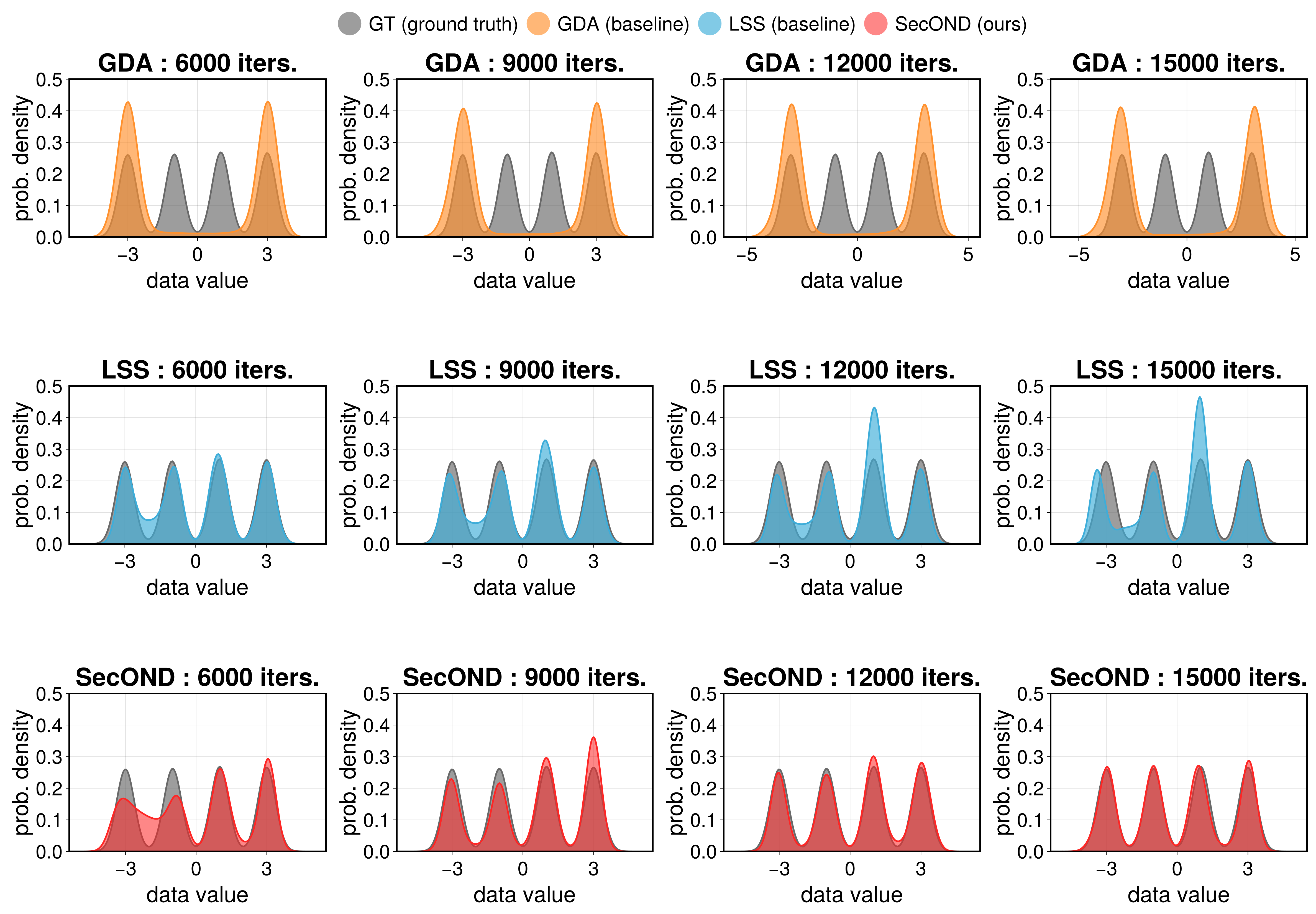}
    \caption{\texttt{SecOND} converges rapidly and to a more accurate solution for a GAN training problem.}
    \label{fig: xinjie gan train ex}
\end{figure}
\subsection{Generative Adversarial Network (GAN)}\label{sec:gan}
Next, we consider a larger-scale test problem in which $\om$ is computed stochastically (i.e., via sampling minibatches of data).
To this end, we evaluated GDA, LSS, and \texttt{SecOND} on a GAN training problem where the generator must fit a 1D mixture of Gaussians with 4 mixture components. The distribution that each algorithm learned at different training iterations is plotted in Figure \ref{fig: xinjie gan train ex}. GDA suffered mode collapse early on and only fit two out of the four modes. Both LSS and \texttt{SecOND} successfully found all four modes of the problem.
While LSS initially seems to converge rapidly, continued training degrades performance.
Over time, \texttt{SecOND} outperformed LSS and fit the ground truth distribution more closely by 12000 iterations.

\subsection{Entropy Regularized Zero-Sum Matrix Games}
 
To test the performance of \texttt{SeCoND} in constrained games with empty interiors, we test it for the following constrained zero-sum game: a regularized matrix game with the following objective:
\begin{equation*}
\begin{split}
    f(\x,\y) =& ~\x^{\top}\y - \underbrace{\mathbf{\left(H(x) - H(y)\right)}}_{\textrm{entropy regularization}}, \quad \x\in \mathbb{R}^2_{+},~\y\in \mathbb{R}^2_{+}, \\
        \mathrm{Player\,1:\,}& \min_{\x \in \mathbb{R}^2_{+}}  ~f(\x, \y) \,\, \quad  \quad \,\,\mathrm{Player\,2:\, } \max_{\y \in \mathbb{R}^2_{+}} ~f(\x, \y), \\
        & \textrm{s.t.}~\x > \mathbf{0},~\y>\mathbf{0}, \mathbf{1}^\top\x = 1~\textrm{and}~\mathbf{1}^\top\y = 1.
\end{split}
\end{equation*}
Here, $\mathbf{H(v)} := \sum_{i=1}^{n} - \mathbf{v}_i \log(\mathbf{v}_i)$ is the entropy function for some $\mathbf{v}\in\mathbb{R}^n_+$. The Nash equilibrium of the above entropy-regularized matrix game is also called the \emph{Quantal Response Equilibrium} (QRE) \citep{mckelvey1998quantal}. Notably, (i) this popular class of games satisfies Assumption \ref{assumption: 3 J top w not 0}, and  (ii) the strategies are constrained to lie in the probability simplex, which has an \emph{empty interior} (but a non-trivial relative interior). 
\begin{figure}
     \centering
     \subcaptionbox{Player 1 $(\x)$ initialization at $[0.1, 0.9]^\top$.}
       {\includegraphics[width=7.5cm]{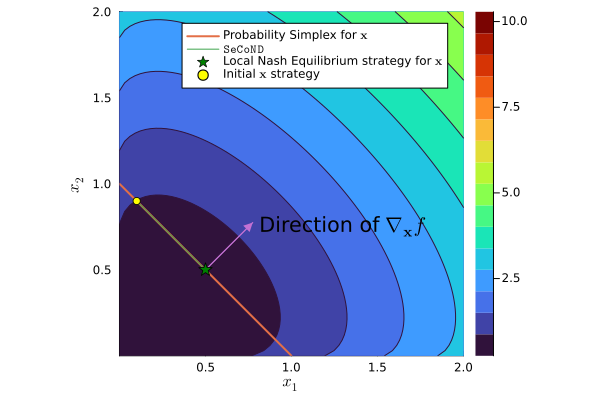}}
     \subcaptionbox{Player 2 $(\y)$ initialization at $[0.9, 0.1]^\top.$}
       {\includegraphics[width=7.5cm]{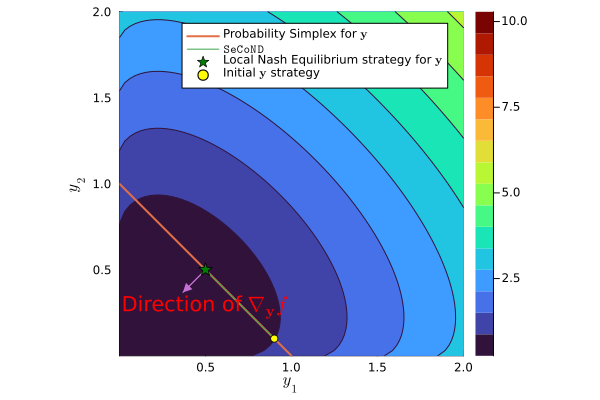}}
     \caption{Numerical results for a constrained game with an empty interior.}\label{fig: qre game}
\end{figure}
\paragraph{Does $\texttt{SeCoND}$ handle cases where the feasible set has an empty interior?} We test \texttt{SeCoND} for the above game, and \Cref{fig: qre game} shows the results for both the players. We observe that \texttt{SeCoND} successfully reaches the local (generalized) Nash equilibrium located at $\x^* = [0.5, 0.5]^\top,~\y^*=[0.5, 0.5]^\top$. 
The projection scheme in Algorithm \ref{algo: constrained} ensures that the updates successfully traverse the simplex, reaching the Nash point. At this Nash point, the direction of $\omega$ is such that $\nabla_{\x} f$ and $\nabla_{\y} f$ are perpendicular to the directions $\x$ and $\y$ can move in their respective probability simplexes, and thus neither player has any incentive to deviate from this point. These results are in line with the convergence guarantee provided by \Cref{theorem: constrained case}.

%% file: sections/06_conc.tex
\section{Conclusion and Future Work}\label{Section: conc}

We provide algorithms that provably converge to only local Nash equilibria in smooth, possibly nonconvex-nonconcave, two-player zero-sum games in the unconstrained (\texttt{DND}, \texttt{SecOND}) and convex-constrained (\texttt{SeCoND}) settings. We show that in the nonconcave-nonconvex setting, \texttt{DND} has an asymptotic \textit{linear} local convergence rate and that \texttt{SecOND} approaches a neighborhood around a fixed point superlinearly. In contrast, the most closely related existing approaches have no established convergence rates in this setting and do not consider constraints. Empirical results demonstrate \texttt{DND} and \texttt{SecOND} outperform previous related works in several test problems. We foresee two main directions for future work. (i) The fundamental links this problem shares with dynamical system theory necessitate second-order information to provide convergence guarantees. Future work should investigate efficiently computable approximations of this second-order information.
(ii) In this paper, we provide guarantees that any point to which the proposed algorithms converge must be a local Nash equilibrium, and conversely that all local Nash equilibria are (local) attractors of these algorithms. These guarantees are in line with other recent works \citep{mazumdar2019finding, adolphs2019local} (even though those works do not provide rates of convergence for the nonconvex-nonconcave regime).
Nevertheless, a key direction of future work in this area will be to establish \emph{global} convergence properties, i.e., to ensure that algorithms converge to local Nash points from any arbitrary initialization (if such a point exists). Currently, in the zero-sum setting, such guarantees only exist for finding local minmax (possibly non-Nash) equilibria in games with restrictive hypercube strategy constraints \citep{daskalakis2023stay}.

%% file: sections/app.tex
\section{Proofs}\label{appendix: proof}

\subsection{Proof of \Cref{lemma: critical points of gc <=> GDA}}
\begin{proof}
    \((\implies) \) Clearly, \begin{equation*}
        \omega(\z)=0\implies g_c(\z) = 0.
    \end{equation*} 
    \((\impliedby)\)
    Now assume that \(\z\) is a critical point of \(g_c\) such that \(\om\neq 0\). In this case, due to the choice of our regularization \(E_c(\z)\), \(g_c(\z)\) can be thought of as \(g_c(\z) = M(\z)J(\z)^\top\om,\text{ where }M(\z)\) is full rank. Thus,
    \begin{equation*}
        g_c(\z) = 0 \implies M(z)J(\z)^\top\om = 0 \implies \om=0, \tag{Assumption \ref{assumption: 3 J top w not 0}}
    \end{equation*}
    which is a contradiction. Hence, \(g_c(\z)=0\iff \omega(\z) = 0\).
\end{proof}

\subsection{Proof of \Cref{theorem: Cont Dyn Sys LNE <=> LASE}}
\begin{proof}
     \((\implies) \) As all LASE points of continuous-time dynamics are also critical points, for any LASE point \(\z = (\x^\top, \y^\top)^\top,\,  \omega(\z) = 0\). Thus the Jacobian of \(g_c\) at \(\z\) becomes
    \begin{equation}\label{eq: Jacobian of gc}
    \begin{split}
    \nabla g_c(\z) &= \left[J(\z)^\top J(\z)(J(\z) + J(\z)^\top)\right]^{-1}J(\z)^\top J(\z) = (J(\z) + J(\z)^\top )^{-1} \\ 
    & = \underbrace{\begin{bmatrix}
             \frac{1}{2}\left(\nabla^2_{\x\x} f(\x, \y)\right)^{-1} & 0 \\ 0 & -\frac{1}{2}\left(\nabla^2_{\y\y} f(\x, \y)\right)^{-1}
         \end{bmatrix}}_{:=H(\z)}.
    \end{split}
    \end{equation}
     From definition \ref{definition: continuous time LASE}, 
     \begin{equation*}
         \nabla g_c(\z) = H(\z) \succ 0 \implies \nabla^2_{\x\x} f(\x, \y) \succ 0 \text{ and }\nabla^2_{\y\y} f(\x, \y) \prec 0,
     \end{equation*}
     which implies that \((\x, \y)\) is a strict local Nash equilibrium of \ref{eq: min x f max y f} (from definition \ref{definition: DNE}). \\
     Thus, every LASE of \(\dot \z = -g_c(\z)\) is a strict local Nash equilibrium of \eqref{eq: min x f max y f}. \\
     \((\impliedby)\) Consider a strict local Nash equilibrium \((\x^*, \y^*)\) of \ref{eq: min x f max y f}. From definition \ref{definition: DNE}, \(\nabla^2_{\x\x} f(\x^*, \y^*) \succ 0, \nabla^2_{\y\y} f(\x^*, \y^*) \prec 0\), and \(\omega(\z^*) = 0\) where \(\z^* = ({\x^*}^\top, {\y^*}^\top)^\top\). Clearly, \(H(\z^*) \succ 0\) and thus \(\z^*\) is a LASE of \eqref{eq: our cont time dyn sys}.
\end{proof}

\subsection{Proof of Corollary \ref{corollaly: symmetric cont case}}
\begin{proof}
    From theorem \ref{theorem: Cont Dyn Sys LNE <=> LASE}, \(\z\) must also be a LASE, and by extension, a critical point of \(g_c\). From lemma \ref{lemma: critical points of gc <=> GDA}, \(\om = 0\). Consider \eqref{eq: Jacobian of gc}. As the inverse Hessians \(\left(\nabla^2_{\x\x} f(\x, \y)\right)^{-1}\) and \(\left(\nabla^2_{\y\y} f(\x, \y)\right)^{-1}\) are symmetric, \(H(\z)\) is symmetric. Because \(\om = 0\), the Jacobian \(\nabla g_c(\z) = H(\z)\), and \(H(\z)\) only has real eigenvalues due to symmetry.
\end{proof}

\subsection{Proof of \Cref{theorem: Discrete time dyn sys <=> LNE}}
\begin{proof}
        \((1. \implies) \) The fixed points of the discrete GDA dynamics in \eqref{eq: GDA discrete dynamical sys} are critical points of \(\omega\), i.e, where \(\om = 0\). Clearly, \begin{equation*}
        \omega(\z)=0\implies g_d(\z) = \z.
    \end{equation*} 
    \((1. \impliedby)\)
    Now assume that \(\z\) is a fixed point of \(g_d\) such that \(\om\neq 0\). In this case, due to the choice of our regularization \(E(\z)\), \(g_d(\z)\) can be thought of as \(g_d(\z) = \z - \alpha M(\z)J(\z)^\top\om,\text{ where }M(\z)\) is full rank and \(\alpha\) is the step size. Thus,
    \begin{equation*}
        g_d(\z) = \z \implies M(z)J(\z)^\top\om = 0 \implies \om=0, \tag{Assumption \ref{assumption: 3 J top w not 0}}
    \end{equation*}
    which is a contradiction. Hence, \(g_d(\z)=\z\iff \omega(\z) = 0\). \\
    \((2. \implies) \) As all LASE points of discrete-time dynamics are also fixed points, for any LASE point \(\z = (\x^\top, \y^\top)^\top,\,  \omega(\z) = 0\). Thus the Jacobian of \(g_d\) at \(\z\) becomes
    \begin{equation}\label{eq:discrete time jacobian}
    \begin{split}
        \nabla g_d(\z) & = I_{n+m} - \alpha(J(\z) + J(\z)^\top + \beta(\z))^{-1}\\
        & = \begin{bmatrix}
            I_n - (2\nabla_{\x\x} f + \mathbbm{1}_{\{\lambda_\x > 0\}}(b_\x)I)^{-1} & 0 \\
            0 & I_m - ( - 2\nabla_{\y\y} f + \mathbbm{1}_{\{\lambda_\y < 0\}}(b_\y)I)^{-1}
        \end{bmatrix}
    \end{split}
    \end{equation}
    The eigenvalues of \(\nabla g_d(\z)\) are the eigenvalues of \(I_n - (2\nabla_{\x\x} f + \mathbbm{1}_{\{\lambda_\x > 0\}}(b_\x)I)^{-1}\) and \(I_m - ( - 2\nabla_{\y\y} f + \mathbbm{1}_{\{\lambda_\y < 0\}}(b_\y)I)^{-1}\).
    For an eigenvalue \(\lambda\) of \(\nabla_{\x\x} f\), the corresponding eigenvalue of \(I_n - (2\nabla_{\x\x} f + \mathbbm{1}_{\{\lambda_\x > 0\}}(b_\x)I)^{-1}\) will be
    \begin{equation}\label{eq: lambda eq for reg}
        1 - \frac{\alpha}{2\lambda + \mathbbm{1}_{\{\lambda_\x > 0\}}(b_\x)}.
    \end{equation}
    If \(\lambda_\x<0\), \Cref{eq: lambda eq for reg} becomes
    \begin{equation}\label{eq: reg makes Nash}
         1 - \frac{\alpha}{2\lambda} > 1. 
    \end{equation}
    As \(\z\) is an LASE point, from Definition \ref{definition: discrete time LASE}, \(\rho(\nabla g_d(\z))<1\).  Thus, \Cref{eq: reg makes Nash} shows that \(\z\) cannot be a LASE if \(\lambda_\x < 0\). Thus \(\z\) is a LASE \(\implies \lambda_\x > 0 \implies \nabla_{\x\x} f \succ 0\). A similar argument by analyzing eigenvalues for \(I_m - ( - 2\nabla_{\y\y} f + \mathbbm{1}_{\{\lambda_\y < 0\}}(b_\y)I)^{-1}\) shows that \(\z\) is a LASE \(\implies \lambda_\y < 0 \implies \nabla_{\y\y} f \prec 0\). Thus, from definition \ref{definition: DNE}, \(z\) is a LASE implies that \(\z\) is a strict local Nash equilibrium of \eqref{eq: min x f max y f}. \\
    \((2. \impliedby)\) Let \(\z\) be a strict local Nash equilibrium. Then, \(\lambda_\x > 0, \lambda_\y < 0\).
    Clearly, from \eqref{eq: reg makes Nash}, all eigenvalues of \(I_n - (2\nabla_{\x\x} f + \mathbbm{1}_{\{\lambda_\x > 0\}}(b_\x)I)^{-1}\) are smaller than 1. Since \(\lambda_\x > 0, \lambda > 0\). Also,\(b_\x > \frac{1}{2}, b_\x > \frac{\alpha}{2}\), which means that
    \begin{equation*}
        1 - \frac{\alpha}{2\lambda + b_\x} > 1 - \frac{\alpha}{2\lambda + \frac{\alpha}{2}} > 1 - \frac{\alpha}{\frac{\alpha}{2}} > -1~\forall~\lambda~\in\textrm{spec}(\nabla_{\x\x}f)
    \end{equation*}
    Thus \(\rho(I_n - (2\nabla_{\x\x} f + \mathbbm{1}_{\{\lambda_\x > 0\}}(b_\x)I)^{-1})<1\). Similarly, \(\rho(I_m - ( - 2\nabla_{\y\y} f + \mathbbm{1}_{\{\lambda_\y < 0\}}(b_\y)I)^{-1})\) is less than 1. Thus, from definition \ref{definition: discrete time LASE}, \(\z\) is also a LASE.\\
    \((3.) \) The Jacobian \(\nabla g_d\) at any fixed point \(\z\) is the same as that given in \Cref{eq:discrete time jacobian}, in which \(\nabla g_d\) is clearly symmetric. Thus, \(\nabla g_d\) only has real eigenvalues at a fixed point \(\z\).
\end{proof}
\subsection{Proof of \Cref{theorem: constrained case}}
\begin{proof}
    Assume that \texttt{SeCoND} converges to a point \(\mathbf{z}\). We consider two cases, as follows:
    \begin{enumerate}
        \item If \(\z\in\mathrm{int\,}\mathcal{G}\), then the immediate neighbourhood around \(\z\) which \texttt{SeCoND} would have to traverse in order to reach \(\z\) is also in \(\mathrm{int\,}\mathcal{G}\). In this neighborhood, the projection step in \texttt{SeCoND} does not have any effect, and the algorithm's dynamics follow \texttt{DND}.
    By theorem \ref{theorem: Discrete time dyn sys <=> LNE}, \texttt{DND} would only have converged to \(\z\) if \(\nabla f (\z) = 0, \nabla_{\x\x}^2 f \succ 0,\) and \(\nabla_{\y\y}^2 f \prec 0,\) which from definition \ref{definition: GNE} implies that \(\z\) is also a strict local Generalized Nash equilibrium.

    \item 
    If \(\z\in\partial\mathcal{G},\) then from Algorithm \ref{algo: constrained}, \(-\om\) must be in the normal cone of $\mathcal{G}$ at \(\z\). Because \(\om = \begin{bmatrix}
        \nabla_\x f \\ - \nabla_\y f
    \end{bmatrix},\) this means that at \(\z\), a feasible step cannot be taken for which \(\x\) or \(\y\) can reduce or increase \(f(\x, \y)\), respectively. 
    Thus, from Definition \ref{definition: GNE}, \(\z\) is a local (not necessarily strict) generalized Nash equilibrium.
    \end{enumerate}

    This concludes the proof.
    \end{proof}
\subsection{Proof of \Cref{theorem: linear convergence rate}}
\begin{proof}
From \Cref{theorem: Discrete time dyn sys <=> LNE}, we know that any non-Nash fixed point is an unstable fixed point of \texttt{DND}. Thus, when \texttt{DND} is randomly initialized (and almost surely not at such an unstable fixed point), such points will certainly be avoided at each iteration; furthermore, if it converges, \texttt{DND} will converge almost surely to a strict local Nash equilibrium \citep{benaim1995dynamics, sastry2013nonlinear}. Let \(\z^*\) denote the strict local Nash equilibrium to which \texttt{DND} converges, and let \(J^\top J (\z)\) denote \(J(\z)^\top J(\z)\). We use Taylor's Theorem \citep[Theorem 2.1]{nocedal1999numerical} applied to \(\omega\), 
\begin{equation*}
    \omega(\z_k) - \omega(\z^*) = \int_0^1 J (\z^* + t(\z_k - \z^*)) (\z_k - \z^*)dt.
\end{equation*}
For large $k$, as $\mathbf{z}_k \rightarrow \mathbf{z}^*$, $J(\mathbf{z}^* + t(\mathbf{z}_k - \mathbf{z}^*)) \approx J(\mathbf{z}_k) \forall t\in[0,1]$. Also for large $k$, from our assumptions $\beta = 0$ and $E = 0$. Thus we get for large k:

\begin{equation*}
\begin{split}
    \lVert\mathbf{z}_{k+1} - \mathbf{z}^*\rVert =& \lVert \mathbf{z}_k - \mathbf{z}^* - \alpha_k [J(\mathbf{z}_k)^\top J(\mathbf{z}_k)(J(\mathbf{z}_k) + J(\mathbf{z}_k)^\top)]^{-1}J(\mathbf{z}_k)^\top \omega(\mathbf{z}_k) \rVert \\
    =& \lVert \mathbf{z}_k - \mathbf{z}^* - \alpha_k(J(\mathbf{z}_k) + J(\mathbf{z}_k)^\top)^{-1} J(\mathbf{z}_k)^{-1} \omega(\mathbf{z}_k)\rVert \\
    =& \lVert \mathbf{z}_k - \mathbf{z}^* - \alpha_k(J(\mathbf{z}_k) + J(\mathbf{z}_k)^\top)^{-1} J(\mathbf{z}_k)^{-1} (\omega(\mathbf{z}_k) - \omega(\mathbf{z}^*))\rVert\\
    =& \lVert \mathbf{z}_k - \mathbf{z}^* - \alpha_k(J(\mathbf{z}_k) + J(\mathbf{z}_k)^\top)^{-1} J(\mathbf{z}_k)^{-1} \left( \int_0^1 J(\mathbf{z}^* + t(\mathbf{z}_k - \mathbf{z}^*)) (\mathbf{z}_k - \mathbf{z}^*) dt \right)\rVert \\
    \approx& \lVert \mathbf{z}_k - \mathbf{z}^* - \alpha_k(J(\mathbf{z}_k) + J(\mathbf{z}_k)^\top)^{-1} J(\mathbf{z}_k)^{-1} J(\mathbf{z}_k) (\mathbf{z}_k - \mathbf{z}^*)\rVert \\
    =& \lVert [I - \alpha_k(J(\mathbf{z}_k) + J(\mathbf{z}_k)^\top)^{-1}](\mathbf{z}_k - \mathbf{z}^*)\rVert \\
    \leq& \lVert I - \alpha_k(J(\mathbf{z}_k) + J(\mathbf{z}_k)^\top)^{-1}\rVert_2 \lVert\mathbf{z}_k - \mathbf{z}^*\rVert
\end{split}
\end{equation*}
Now, consider the matrix $D_k = I - \alpha_k(J(\mathbf{z}_k) + J(\mathbf{z}_k)^\top)^{-1}$. From the structure of \(J(\z_k)\) described in \eqref{eq: structure of J},
\begin{equation*}
    D_k = \begin{bmatrix}
        I - \frac{\alpha_k}{2}(\nabla_{\x\x})^{-1} & 0 \\ 0 &   I + \frac{\alpha_k}{2}(\nabla_{\y\y})^{-1}
    \end{bmatrix}.
\end{equation*}
From the properties of \(\lVert\cdot\rVert_2\) norm,
\begin{equation*}
    \lVert D_k \rVert_2 = \max \left\{ \lVert I - \frac{\alpha_k}{2}(\nabla_{\x\x})^{-1}\rVert_2,  \lVert I + \frac{\alpha_k}{2}(\nabla_{\y\y})^{-1}\rVert_2\right\}
\end{equation*}
Let \(\lambda_\x, \lambda_y\) denote the quantities in \eqref{eq: choosing beta for disc dyn sys}, evaluated at \(\z = \z_k\). Further, let \(\tilde{\lambda}_\x, \tilde{\lambda}_y\) denote \(\lambda_\x, \lambda_y\) evaluated at \(\lim_{k\rightarrow\infty}\z_k\). Then, from theorem \ref{theorem: Discrete time dyn sys <=> LNE}, \(\tilde{\lambda}_\x>0, \tilde{\lambda}_y<0\). Thus we can write
\begin{equation*}
    \lim_{k\rightarrow\infty} \lVert D_k \rVert_2 = \max \left\{1 - \frac{\alpha}{2\tilde{\lambda}_\x}, 1 + \frac{\alpha}{2 \tilde{\lambda}_y}\right\} <  1~\forall ~0 < \alpha \leq \max\{2| \tilde{\lambda}_x|, 2| \tilde{\lambda}_y|\}
\end{equation*}
Thus, 
\begin{equation*}
    \lim_{k\rightarrow\infty} \frac{\lVert\z_{k+1} - \z^*\rVert}{\lVert\z_k - \z^*\rVert} \leq \lim_{k\rightarrow\infty}\lVert D_k \rVert_2 < 1
\end{equation*}
This proves that \texttt{DND} has a local linear convergence rate when the step size is chosen as described.

\end{proof}

\subsection{Proof of \Cref{theorem: modified stable}}
\begin{proof}
    First, we show that the fixed points of \texttt{SecOND} and \texttt{DND} are the same. 
    From \eqref{eq: our modified dis dyn sys}, any fixed point \(\z\) of \texttt{SecOND} must have \(\om = 0\), i.e., fixed points \(\z\) of algorithm \ref{algo: modified discrete} are same as the fixed points of the discrete-time GDA dynamics. Theorem \ref{theorem: Discrete time dyn sys <=> LNE} has already established that the fixed points of the discrete GDA dynamics are the same as the fixed points of \texttt{DND}.

    From \eqref{eq: our modified dis dyn sys}, when far away from \(\z_c\), \texttt{SecOND} satisfies the condition that every step is in a feasible descent direction. Further, using a line search rule like \eqref{eq: armijo} ensures that for every step that \second~takes far away from \(\z_c\), the merit function \(\|\om\|^2_2\) decreases in value. Thus, when \(S(\z_k) \approx J(\z_k)^\top J(\z_k),\) \texttt{SecOND} mimics a Gauss-Newton method and from standard nonlinear programming results \citep[Proposition 1.1.4]{bertsekas1997nonlinear}, reaches the neighborhood of \(\z_c\) superlinearly. Now, when \second~reaches this neighborhood, it changes its dynamics to \texttt{DND}, which has already shown to have only local Nash equilibrium points as its LASE points. Clearly, \second~has the same LASE points as \texttt{DND} once it switches dynamics, and results from theorem \ref{theorem: Discrete time dyn sys <=> LNE} apply and \second~only converges to a strict local Nash equilibrium. Let us derive the local superlinear rate now. Let \(\mathcal{B}_{\delta}(\z_c)\) denote a ball of radius \(\delta\) centered at \(\z_c\), and assume that \(\z_0\in\mathcal{B}_{\delta}(\z_c)\). Let \(S(\z_k)\) be denoted by \(S_k\). For iteration \(k\) when \(\lVert\z_k - \z_{k-1}\rVert > \epsilon\):
    \begin{equation*}
    \begin{split}
        \|\z_{k+1} - \z_c\| =& \|\z_k - S_k^{-1}J(\z_k)^\top \omega(\z_k) - \z_c\| \\
        =& \lVert S_k^{-1} (S_k(\z_k - \z_c) - J(\z_k)^\top \omega(\z_k)) \rVert\\
        =& \lVert S_k^{-1} \left(S_k - J(\z_k)^\top \int^1_0 J(\z_c + t(\z_k - \z_c))dt \right)(\z_k-\z_c)\rVert\\
        =& \lVert S_k^{-1} \left(\int^1_0 \left[ S_k - J(\z_k)^\top J(\z_c + t(\z_k - \z_c))\right]dt \right)(\z_k-\z_c)\rVert\\
        \leq & \lVert S_k^{-1}\rVert \lVert \left(\int^1_0 \left[ S_k - J(\z_k)^\top J(\z_c + t(\z_k - \z_c))\right]dt \right)\rVert \lVert(\z_k-\z_c)\rVert
    \end{split}
    \end{equation*}
    By choosing \(S_k = J(\z_k)^\top J(\z_k)\), we get
    \begin{align*}
        \|\z_{k+1} - \z_c\| \leq& \lVert S_k^{-1}\rVert \lVert \left(\int^1_0 \left[ J(\zk)^\top \left(J(\zk) - J(\z_c + t(\z_k - \z_c))\right)\right]dt \right)\rVert \lVert(\z_k-\z_c)\rVert\\
        \leq& \lVert S_k^{-1}\rVert\lVert J(\zk)^\top\rVert \lVert \left(\int^1_0   \|J(\zk) - J(\z_c + t(\z_k - \z_c))\|dt \right)\rVert \lVert(\z_k-\z_c)\rVert\\
         \leq& ML_\omega L_J  \left(\int^1_0  (1-t)dt \right)\lVert\z_k-\z_c\rVert^2 = \frac{ML_\omega L_J}{2}\|\zk-\z_c\|^2
    \end{align*}
    Similarly to \texttt{DND} in the proof of \Cref{theorem: linear convergence rate}, any non-Nash fixed point of \texttt{SecOND} is unstable. Thus when \texttt{SecOND} is randomly initialized (and almost surely not at such an unstable fixed point), such unstable fixed points will be certainly avoided \citep{benaim1995dynamics, sastry2013nonlinear}, and thus if it converges, \texttt{SecOND} converges almost surely to a strict local Nash equilibrium.
\end{proof}

\section{Relaxing Assumption \ref{assumption: 3 J top w not 0}}\label{appendix: note on assumptions}
In this section, we show that simple modifications to our various dynamics eliminate the need for Assumption \ref{assumption: 3 J top w not 0}, without influencing convergence guarantees, by adding a time-varying term to the dynamics. We illustrate this through our continuous-time dynamical system, $\dot \z = -g_c(\z)$ from \Cref{eq: our cont time dyn sys}. Adding a time varying term gives rise to a system of the form $\dot \z = -g_c(\z, t)$. We first mention the stability conditions of an equilibrium for such a system.
\begin{definition}[LASE of a Continuous Time-varying System, \citep{sastry2013nonlinear}]\label{definition: cont time varying lase}
    A \\ point $\z^*\in\mathbb{R}^{n+m}$ is a LASE of the continuous time-varying dynamical system $\dot \z = - h_c(\z, t)$ if $h_c(\z^*, t) = 0~\forall~t\geq 0$, and $\lambda>0~\forall~\lambda\in\textrm{spec}\left(\nabla_\z h_c(\z^*, t) + \nabla_\z h_c(\z^*, t)^\top\right)~\forall ~t\geq 0$.
\end{definition}

We choose a time varying term $h(\z,t)$ to construct a desirable modified system $\dot \z = -g_c(\z) + h(\z, t)$. We show that a choice made in existing zero-sum games literature can be readily adapted to relax Assumption \ref{assumption: 3 J top w not 0} in our case \citep{mazumdar2019finding}.
 Consider $h(\z, t) = a(1-e^{-b\|\om\|^2})e^{-t}\tilde{\z}$, for an arbitrary constant vector $\tilde{\z}\in\mathbb{R}^{n+m}\setminus\{0\}$, and $a,b>0$ are positive scalars. Then, we have that: $ h(\z,t)$ is bounded, $\om = 0 \iff h(\z,t) = 0$, and $\om = 0 \implies \nabla_\z h(\z,t) = 0$.

Following Definition \ref{definition: cont time varying lase}, this choice of $h(\z, t)$ implies that any LASE of $\dot \z = -g_c(\z, t)$ is also an LASE of $\dot z = -g_c(\z)$, and thus a local Nash equilibrium of \ref{eq: min x f max y f}. Further, any local Nash equilibrium of \ref{eq: min x f max y f} is also an LASE of $\dot \z = -g_c(\z, t)$. Thus, we get a desirable analog of \Cref{theorem: Cont Dyn Sys LNE <=> LASE} without the need for Assumption \ref{assumption: 3 J top w not 0}. One may conduct a similar analysis for the discrete-time dynamics presented in our paper by adding a time varying term as for the continuous case.

\section{Choice of Regularization Matrices and S in \texttt{SecOND}}\label{appendix: choice of S in SecOND}
\paragraph{Choice of S(z) for superlinear Gauss-Newton Interpretation.}
We take \(S(\z_k)\) to be $J(\mathbf{z}_k)^\top J(\mathbf{z}_k) + \lambda_k I$ where \(\lim_{k\to\infty}\lambda_k = 0\).\par
A way of designing regularization matrices is by using the Gershgorin Circle Theorem \citep{horn2012matrix}, which states that for a matrix $A \in \mathbb{C}^{n\times n} $, all eigenvalues of $A$ lie in the union of $n$ discs centred at $A_{ii}$ with radii $R_i = \sum_{j = 1, j\neq i}^{j=n} |A_{ij}|$ for $i=1,\dots,n$. Thus, to regularize $A$ for invertibility, a diagonal regularization matrix $M$ with the $i^{th}$ diagonal entry $M_{ii} = \mathbbm{1}_{\{A_{ii}-R_i < 0\}}(|A_{ii}-R_i| + \lambda_0)$, where $\lambda_0>0$ is user specified and is a lower bound on the real part of eigenvalues of $A+M$. With this, we specify:

\begin{enumerate}
    \item \textbf{Design of $E_c(\mathbf{z})$ in \eqref{eq: our cont time dyn sys}:} Here, $A = J(\mathbf{z})^\top J(\mathbf{z})(J(\mathbf{z})+J(\mathbf{z})^\top)$, and the regularization matrix \newline $E_{c}(\mathbf{z})_{ii} = \mathbbm{1}_{\{A_{ii}-R_i < 0\text{ and }\lVert\omega(\mathbf{z})\rVert>\delta_0\}}(|A_{ii}-R_i| + \lambda_0)$. The constant $\delta_0 > 0$ is also user-specified and ensures that at a critical point, $E_c$ is differentiable and that $E_c = 0$.
    \item \textbf{Design of $E(\mathbf{z}_k)$ in \eqref{eq: our dis dyn sys}:} In this case, $A = J(\mathbf{z}_k)^\top J(\mathbf{z}_k)(J(\mathbf{z}_k)+J(\mathbf{z}_k)^\top + \beta(\mathbf{z}_k))$, and we proceed as above.
    \item \textbf{Design of $S(\mathbf{z}_k)$ in \eqref{eq: our modified dis dyn sys}:} We can take $A$ as the matrix given in Equation (18) and choose $\lambda = \max_{i} \{(A_{ii} - R_i) + \lambda_0\}$ (and thus $S = A + \lambda I$). For the Gauss-Newton interpretation, we can take $A=J(\mathbf{z}_k)^\top J(\mathbf{z}_k)$.
\end{enumerate}  In our experiments, we took the values $\lambda_0 = 5$ and $\delta_0$ = $5\times 10^{-5}$.

\section{Experimental Details}\label{Appendix: Exp details}
\subsection{Two-Dimensional Toy Example}
\subsubsection{Baselines}\label{appendix: toy baselines}
\begin{figure}
    \centering
    \includegraphics[width=7.5cm]{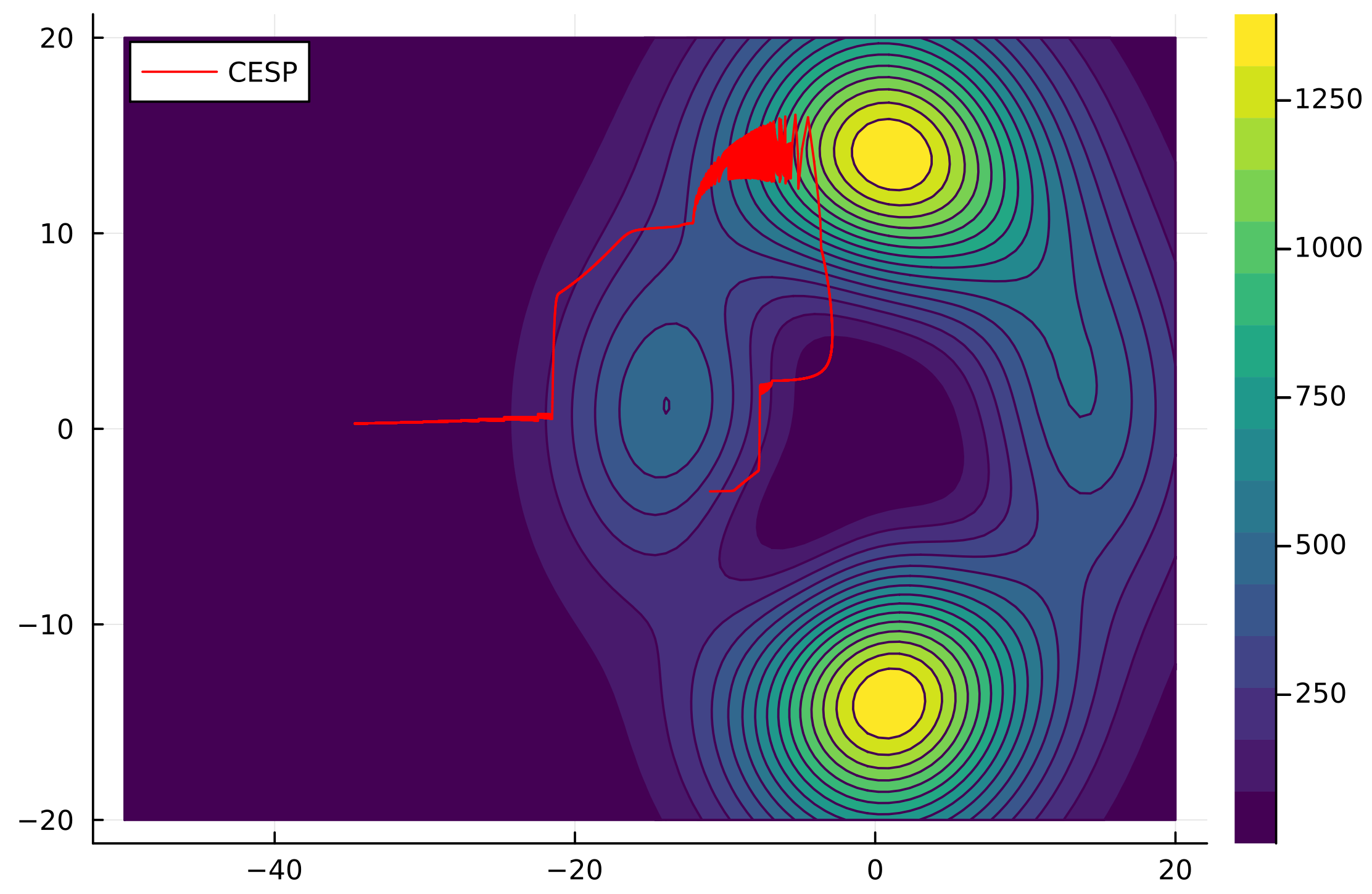}
    \caption{CESP \citep{adolphs2019local} diverges for the two-dimensional toy example.}
    \label{fig:cesp}
\end{figure}

\paragraph{Local Symplectic Surgery (LSS).} For the toy example, the LSS method is:
\begin{equation*}
    \z_{k+1} = \z_k - \alpha (\omega(\z_k) + e^{-\xi_2 \|v\|^2}v),
\end{equation*}
where \(v = J(\z_k)^\top (J(\z_k)^\top J(\z_k) + \lambda(\z_k) I)^{-1} J(\z_k)^\top \omega(\z_k)\) and regularization \(\lambda(\z_k) = \xi_1(1-e^{\|\omega(\z_k)\|^2})\). Here, \(\xi_1 = \xi_2 = 10^{-4}\). These values have been recommended in the LSS paper for this particular example. Though the authors also described a two-timescale discrete system of LSS, it could not reliably converge for this example, and thus, we resorted to the equation above.
\paragraph{(Curvature Exploitation for the
Saddle Point problem (CESP).} The CESP method is given by:
\begin{equation*}
    \z_{k+1} = \z_k - \alpha \omega(\z_k) + \begin{bmatrix}
        \mathbf{v}_{\z_k}^{(-)} \\ \mathbf{v}_{\z_k}^{(+)}
    \end{bmatrix},
\end{equation*}
where, for the sign function \(\mathrm{sgn:}\mathbb{R}\rightarrow\{-1,1\}\),
\begin{equation*}
    \begin{split}
        \mathbf{v}_{\z_k}^{(-)} &= \mathbbm{1}_{\lambda_\x<0}\frac{\lambda_\x}{2\rho_\x}\mathrm{sgn}(\mathbf{v}_\x^\top \nabla_\x f(\x, \y))\mathbf{v}_\x \\
        \mathbf{v}_{\z_k}^{(+)} &= \mathbbm{1}_{\lambda_\y>0}\frac{\lambda_\y}{2\rho_\y}\mathrm{sgn}(\mathbf{v}_\y^\top \nabla_\y f(\x, \y))\mathbf{v}_\y.
    \end{split}
\end{equation*}
Here, \(\lambda_\x\) and \(\lambda_\y\) denote the minimum and maximum eigenvalues of \(\nabla^2_{\x\x} f\) and \(\nabla^2_{\y\y} f\) respectively. \(\mathbf{v}_\x\) and \(\mathbf{v}_\y\) denote the eigenvectors of \(\lambda_\x\) and \(\lambda_\y\). We took \(\nicefrac{1}{2\rho_\x} = \nicefrac{1}{2\rho_\y} = 0.05\). CESP could not converge reliably for the two-dimensional example, and a typical diverging plot is shown in figure \ref{fig:cesp}.
\subsubsection{Experiment Parameters.} \label{appendix: toy ex params}
For all algorithms, step size \(\alpha\) was taken to be 0.001, except for \texttt{SecOND} which performed Armijo line search. Tolerance for convergence was set at \(10^{-5}\), and the maximum number of allowable iterations for every algorithm was 15,000. \(\epsilon\) for \texttt{SecOND} (algorithm \ref{algo: modified discrete}) was taken to be \(10^{-2}\). For figure \ref{fig: Unconstrained Toy}, data points that were below \(Q_1 -1.5(Q_3-Q_1)\) or above \(Q_3 + 1.5(Q_3-Q_1)\) were considered outliers. Here, \(Q_1\) and \(Q_3\) denote the first and third quartiles, respectively.
\subsubsection{Additional Unconstrained Case Result.}\label{appendix: additional result}
\begin{figure}
    \centering
    \includegraphics[width=7.5cm]{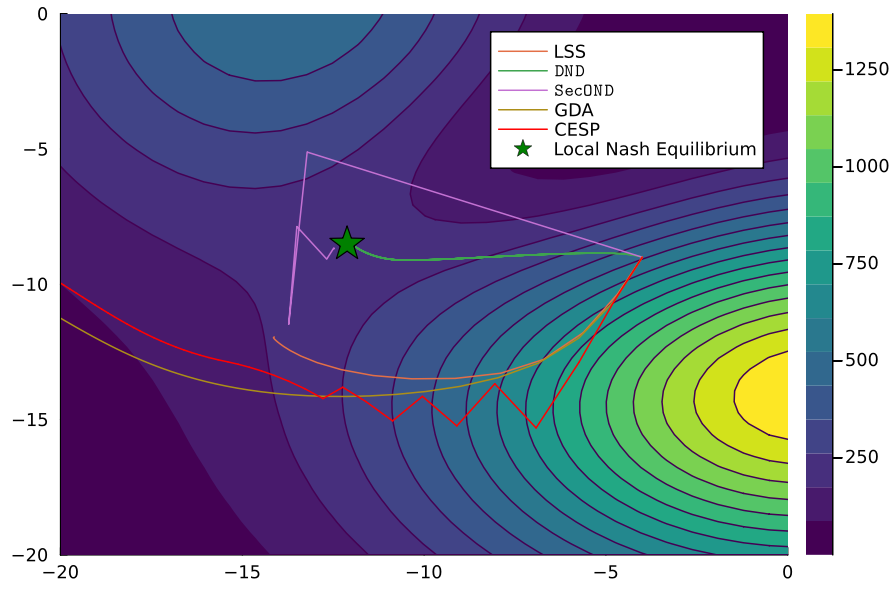}
    \caption{\texttt{SecOND} and \texttt{DND} converge successfully to a local Nash equilibrium.}
    \label{fig: additional result}
\end{figure}

We show a comparison of \texttt{SecOND} and \texttt{DND} for the unconstrained toy example to show that our approaches converge to local Nash equilibrium. From figure \ref{fig: additional result}, it can be seen that only \texttt{SecOND} and \texttt{DND} successfully converge to local Nash equilibrium. CESP and GDA diverged, while LSS converged to a non-Nash point. This behavior of LSS arises due to the assumption they make (Theorem 4, \citep{mazumdar2019finding}), which gets violated. Out of the algorithms which converged, LSS took 75 iterations, \text{DND} took 5405 iterations, while \texttt{SecOND} took  just 7 iterations.
\subsection{Generative Adversarial Network}
\subsubsection{LSS Baseline.}
For GAN training, we use the two-timescale approximation method for LSS described in \citep{mazumdar2019finding}, which is given by
\begin{equation*}
    \begin{split}
        \z_{k+1} =& \z_k - \gamma_1(\omega(\z_k) + e^{-\xi_2\|J(\z_k)^\top v_k\|^2}J(\z_k)^\top v_k) \\
        v_{k+1} =& v_k -\gamma_2(J(\z_k)^\top J(\z_k)v_k + \lambda(\z_n)v_k - J(\z_k)^\top \omega(\z_k)).
    \end{split}
\end{equation*}
Similar to the toy example, \(\lambda(\z_k) = \xi_1(1-e^{\|\omega(\z_k)\|^2})\), and \(\xi_1=\xi_2 = 10^{-4}\)
In the GAN example in Section \ref{sec:gan}, the zero-sum game is between the generator~$G$, which minimizes \(\mathcal{F}\), and the discriminator~$D$, which maximizes \(\mathcal{F}\). Here, \(\mathcal{F}:= \mathbb{E}_{x \sim p_{\mathrm{data}}(x)}[\log D(x)] + \mathbb{E}_{\epsilon \sim p_{\epsilon}(\epsilon)}[\log(1 - D(G(\epsilon)))]\), and \(x\) and \(\epsilon\) denote actual data samples and noise samples, respectively. Table \ref{tab:model_params} lists the parameter values of the GAN model used in our evaluation. 

\begin{table}[h]
\centering
\caption{Parameters of the GAN example in Section \ref{sec:gan}.}
\label{tab:model_params}
\begin{tabular}{l|r|r}

                     & Discriminator & Generator \\ \hline \hline
Input Dimension      & 1                               & 1                            \\\hline
Hidden Layers        & 2                                 & 2                             \\\hline
Hidden Units / Layer & 8                                & 8                            \\\hline
Activation Function  & tanh                              & tanh                          \\\hline
Output Dimension     & 1                                 & 1                           \\\hline
Batch Size           & \multicolumn{2}{c}{128}                                          \\\hline
Dataset size         & \multicolumn{2}{c}{10000}                                           \\\hline
\end{tabular}
\end{table}

We evaluate GDA, LSS, and our \texttt{SecOND} approach. GDA uses an Adam optimizer with a learning rate~$10^{-4}$; LSS uses an RMSProp optimizer with a learning rate~$2 \times 10^{-4}$ for the~$x$ and~$y$ processes and~$1 \times 10^{-5}$ for the~$v$ process, as reported in~\citep{mazumdar2019finding}. \texttt{SecOND} uses an RMSProp optimizer with a learning rate~$2 \times 10^{-4}$. 

\paragraph{Remark} As suggested by \citep{goodfellow2014generative}, to improve the convergence of GDA, we update the discriminator~$k = 3$ times more frequent than the generator~$G$. Moreover, the GDA generator maximizes~$\log(D(G(\epsilon)))$ instead of minimizing~$\log(1 - D(G(\epsilon)))$. We found the best practical performance with the said setup. 

\section{Hardware}\label{appendix: hardware}
The two-dimensional toy and entropy-regularized matrix game examples were run on an Intel i7-11800H 8-core CPU. 
The GAN training sessions were run on an AMD Ryzen 9 7950X 16-core CPU.

%% file: sample.bbl
\begin{thebibliography}{57}
\providecommand{\natexlab}[1]{#1}
\providecommand{\url}[1]{\texttt{#1}}
\expandafter\ifx\csname urlstyle\endcsname\relax
  \providecommand{\doi}[1]{doi: #1}\else
  \providecommand{\doi}{doi: \begingroup \urlstyle{rm}\Url}\fi

\bibitem[Abernethy et~al.(2021)Abernethy, Lai, and Wibisono]{abernethy2021last}
Jacob Abernethy, Kevin~A Lai, and Andre Wibisono.
\newblock Last-iterate convergence rates for min-max optimization: Convergence of hamiltonian gradient descent and consensus optimization.
\newblock In \emph{Algorithmic Learning Theory}, pages 3--47. PMLR, 2021.

\bibitem[Adolphs et~al.(2019)Adolphs, Daneshmand, Lucchi, and Hofmann]{adolphs2019local}
Leonard Adolphs, Hadi Daneshmand, Aurelien Lucchi, and Thomas Hofmann.
\newblock Local saddle point optimization: A curvature exploitation approach.
\newblock In \emph{The 22nd International Conference on Artificial Intelligence and Statistics}, pages 486--495. PMLR, 2019.

\bibitem[Armijo(1966)]{armijo1966minimization}
Larry Armijo.
\newblock Minimization of functions having lipschitz continuous first partial derivatives.
\newblock \emph{Pacific Journal of mathematics}, 16\penalty0 (1):\penalty0 1--3, 1966.

\bibitem[Attias et~al.(2025)Attias, Dagan, Daskalakis, Yao, and Zampetakis]{attias2025fixed}
Idan Attias, Yuval Dagan, Constantinos Daskalakis, Rui Yao, and Manolis Zampetakis.
\newblock Fixed point computation: Beating brute force with smoothed analysis.
\newblock \emph{arXiv preprint arXiv:2501.10884}, 2025.

\bibitem[Azizian et~al.(2020)Azizian, Mitliagkas, Lacoste-Julien, and Gidel]{azizian2020tight}
Wa{\"\i}ss Azizian, Ioannis Mitliagkas, Simon Lacoste-Julien, and Gauthier Gidel.
\newblock A tight and unified analysis of gradient-based methods for a whole spectrum of differentiable games.
\newblock In \emph{International conference on artificial intelligence and statistics}, pages 2863--2873. PMLR, 2020.

\bibitem[Azizian et~al.(2024)Azizian, Iutzeler, Malick, and Mertikopoulos]{Azizian2024}
Wa\"{\i}ss Azizian, Franck Iutzeler, J\'{e}r\^{o}me Malick, and Panayotis Mertikopoulos.
\newblock The rate of convergence of bregman proximal methods: Local geometry versus regularity versus sharpness.
\newblock \emph{SIAM Journal on Optimization}, 34\penalty0 (3):\penalty0 2440--2471, 2024.
\newblock \doi{10.1137/23M1580218}.
\newblock URL \url{https://doi.org/10.1137/23M1580218}.

\bibitem[Balduzzi et~al.(2018)Balduzzi, Racaniere, Martens, Foerster, Tuyls, and Graepel]{balduzzi2018mechanics}
David Balduzzi, Sebastien Racaniere, James Martens, Jakob Foerster, Karl Tuyls, and Thore Graepel.
\newblock The mechanics of n-player differentiable games.
\newblock In \emph{International Conference on Machine Learning}, pages 354--363. PMLR, 2018.

\bibitem[Ben-Tal et~al.(2009)Ben-Tal, El~Ghaoui, and Nemirovski]{ben2009robust}
Aharon Ben-Tal, Laurent El~Ghaoui, and Arkadi Nemirovski.
\newblock \emph{Robust Optimization}, volume~28.
\newblock Princeton university press, 2009.

\bibitem[Bena{\"\i}m and Hirsch(1995)]{benaim1995dynamics}
Michel Bena{\"\i}m and Morris~W Hirsch.
\newblock Dynamics of morse-smale urn processes.
\newblock \emph{Ergodic Theory and Dynamical Systems}, 15\penalty0 (6):\penalty0 1005--1030, 1995.

\bibitem[Bena{\"\i}m and Hirsch(1999)]{benaim1999mixed}
Michel Bena{\"\i}m and Morris~W Hirsch.
\newblock Mixed equilibria and dynamical systems arising from fictitious play in perturbed games.
\newblock \emph{Games and Economic Behavior}, 29\penalty0 (1-2):\penalty0 36--72, 1999.

\bibitem[Bertsekas(1997)]{bertsekas1997nonlinear}
Dimitri~P Bertsekas.
\newblock Nonlinear programming.
\newblock \emph{Journal of the Operational Research Society}, 48\penalty0 (3):\penalty0 334--334, 1997.

\bibitem[Bo{\c{t}} and B{\"o}hm(2023)]{boct2023alternating}
Radu~Ioan Bo{\c{t}} and Axel B{\"o}hm.
\newblock Alternating proximal-gradient steps for (stochastic) nonconvex-concave minimax problems.
\newblock \emph{SIAM Journal on Optimization}, 33\penalty0 (3):\penalty0 1884--1913, 2023.

\bibitem[Cai et~al.(2024)Cai, Oikonomou, and Zheng]{cai2024accelerated}
Yang Cai, Argyris Oikonomou, and Weiqiang Zheng.
\newblock Accelerated algorithms for constrained nonconvex-nonconcave min-max optimization and comonotone inclusion.
\newblock In \emph{International Conference on Machine Learning}, pages 5312--5347. PMLR, 2024.

\bibitem[Chinchilla et~al.(2023)Chinchilla, Yang, and Hespanha]{chinchilla2023newton}
Raphael Chinchilla, Guosong Yang, and Jo{\~a}o~P Hespanha.
\newblock Newton and interior-point methods for (constrained) nonconvex--nonconcave minmax optimization with stability and instability guarantees.
\newblock \emph{Mathematics of Control, Signals, and Systems}, pages 1--41, 2023.

\bibitem[Daskalakis et~al.(2017)Daskalakis, Ilyas, Syrgkanis, and Zeng]{daskalakis2017training}
Constantinos Daskalakis, Andrew Ilyas, Vasilis Syrgkanis, and Haoyang Zeng.
\newblock Training gans with optimism.
\newblock \emph{arXiv preprint arXiv:1711.00141}, 2017.

\bibitem[Daskalakis et~al.(2021)Daskalakis, Skoulakis, and Zampetakis]{daskalakis2021complexity}
Constantinos Daskalakis, Stratis Skoulakis, and Manolis Zampetakis.
\newblock The complexity of constrained min-max optimization.
\newblock In \emph{Proceedings of the 53rd Annual ACM SIGACT Symposium on Theory of Computing}, pages 1466--1478, 2021.

\bibitem[Daskalakis et~al.(2023)Daskalakis, Golowich, Skoulakis, and Zampetakis]{daskalakis2023stay}
Constantinos Daskalakis, Noah Golowich, Stratis Skoulakis, and Emmanouil Zampetakis.
\newblock Stay-on-the-ridge: Guaranteed convergence to local minimax equilibrium in nonconvex-nonconcave games.
\newblock In \emph{The Thirty Sixth Annual Conference on Learning Theory}, pages 5146--5198. PMLR, 2023.

\bibitem[Diakonikolas et~al.(2021)Diakonikolas, Daskalakis, and Jordan]{diakonikolas2021efficient}
Jelena Diakonikolas, Constantinos Daskalakis, and Michael~I Jordan.
\newblock Efficient methods for structured nonconvex-nonconcave min-max optimization.
\newblock In \emph{International Conference on Artificial Intelligence and Statistics}, pages 2746--2754. PMLR, 2021.

\bibitem[Facchinei and Kanzow(2010{\natexlab{a}})]{facchinei2010generalized}
Francisco Facchinei and Christian Kanzow.
\newblock Generalized nash equilibrium problems.
\newblock \emph{Annals of Operations Research}, 175\penalty0 (1):\penalty0 177--211, 2010{\natexlab{a}}.

\bibitem[Facchinei and Kanzow(2010{\natexlab{b}})]{facchinei2010penalty}
Francisco Facchinei and Christian Kanzow.
\newblock Penalty methods for the solution of generalized nash equilibrium problems.
\newblock \emph{SIAM Journal on Optimization}, 20\penalty0 (5):\penalty0 2228--2253, 2010{\natexlab{b}}.

\bibitem[Facchinei and Pang(2003)]{facchinei2003finite}
Francisco Facchinei and Jong-Shi Pang.
\newblock \emph{Finite-dimensional Variational Inequalities and Complementarity Problems}.
\newblock Springer, 2003.

\bibitem[Fiez and Ratliff(2021)]{fiez2021local}
Tanner Fiez and Lillian~J Ratliff.
\newblock Local convergence analysis of gradient descent ascent with finite timescale separation.
\newblock In \emph{Proceedings of the International Conference on Learning Representations (ICLR)}, 2021.

\bibitem[Fiez et~al.(2020)Fiez, Chasnov, and Ratliff]{fiez2020implicit}
Tanner Fiez, Benjamin Chasnov, and Lillian Ratliff.
\newblock Implicit learning dynamics in stackelberg games: Equilibria characterization, convergence analysis, and empirical study.
\newblock In \emph{International Conference on Machine Learning}, pages 3133--3144. PMLR, 2020.

\bibitem[Gidel et~al.(2018)Gidel, Berard, Vignoud, Vincent, and Lacoste-Julien]{gidel2018variational}
Gauthier Gidel, Hugo Berard, Ga{\"e}tan Vignoud, Pascal Vincent, and Simon Lacoste-Julien.
\newblock A variational inequality perspective on generative adversarial networks.
\newblock \emph{arXiv preprint arXiv:1802.10551}, 2018.

\bibitem[Gidel et~al.(2019)Gidel, Hemmat, Pezeshki, Le~Priol, Huang, Lacoste-Julien, and Mitliagkas]{gidel2019negative}
Gauthier Gidel, Reyhane~Askari Hemmat, Mohammad Pezeshki, R{\'e}mi Le~Priol, Gabriel Huang, Simon Lacoste-Julien, and Ioannis Mitliagkas.
\newblock Negative momentum for improved game dynamics.
\newblock In \emph{The 22nd International Conference on Artificial Intelligence and Statistics}, pages 1802--1811. PMLR, 2019.

\bibitem[Goodfellow et~al.(2014)Goodfellow, Pouget-Abadie, Mirza, Xu, Warde-Farley, Ozair, Courville, and Bengio]{goodfellow2014generative}
Ian Goodfellow, Jean Pouget-Abadie, Mehdi Mirza, Bing Xu, David Warde-Farley, Sherjil Ozair, Aaron Courville, and Yoshua Bengio.
\newblock Generative adversarial nets.
\newblock \emph{Advances in neural information processing systems}, 27, 2014.

\bibitem[Grimmer et~al.(2023)Grimmer, Lu, Worah, and Mirrokni]{grimmer2023landscape}
Benjamin Grimmer, Haihao Lu, Pratik Worah, and Vahab Mirrokni.
\newblock The landscape of the proximal point method for nonconvex--nonconcave minimax optimization.
\newblock \emph{Mathematical Programming}, 201\penalty0 (1):\penalty0 373--407, 2023.

\bibitem[Grnarova et~al.(2017)Grnarova, Levy, Lucchi, Hofmann, and Krause]{grnarova2017online}
Paulina Grnarova, Kfir~Y Levy, Aurelien Lucchi, Thomas Hofmann, and Andreas Krause.
\newblock An online learning approach to generative adversarial networks.
\newblock \emph{arXiv preprint arXiv:1706.03269}, 2017.

\bibitem[Hommes and Ochea(2012)]{hommes2012multiple}
Cars~H Hommes and Marius~I Ochea.
\newblock Multiple equilibria and limit cycles in evolutionary games with logit dynamics.
\newblock \emph{Games and Economic Behavior}, 74\penalty0 (1):\penalty0 434--441, 2012.

\bibitem[Horn and Johnson(2012)]{horn2012matrix}
Roger~A Horn and Charles~R Johnson.
\newblock \emph{Matrix analysis}.
\newblock Cambridge university press, 2012.

\bibitem[Isaacs(1999)]{isaacs1999differential}
Rufus Isaacs.
\newblock \emph{Differential Games: a Mathematical Theory with Applications to Warfare and Pursuit, Control and Optimization}.
\newblock Courier Corporation, 1999.

\bibitem[Jin et~al.(2020)Jin, Netrapalli, and Jordan]{jin2020local}
Chi Jin, Praneeth Netrapalli, and Michael Jordan.
\newblock What is local optimality in nonconvex-nonconcave minimax optimization?
\newblock In \emph{International conference on machine learning}, pages 4880--4889. PMLR, 2020.

\bibitem[Lanczos(1950)]{lanczos1950iteration}
Cornelius Lanczos.
\newblock An iteration method for the solution of the eigenvalue problem of linear differential and integral operators.
\newblock \emph{Journal of Research of the National Bureau of Standards}, 45:\penalty0 255--282, 1950.

\bibitem[Liang and Stokes(2019)]{liang2019interaction}
Tengyuan Liang and James Stokes.
\newblock Interaction matters: A note on non-asymptotic local convergence of generative adversarial networks.
\newblock In \emph{The 22nd International Conference on Artificial Intelligence and Statistics}, pages 907--915. PMLR, 2019.

\bibitem[Lin et~al.(2020)Lin, Jin, and Jordan]{lin2020gradient}
Tianyi Lin, Chi Jin, and Michael Jordan.
\newblock On gradient descent ascent for nonconvex-concave minimax problems.
\newblock In \emph{International conference on machine learning}, pages 6083--6093. PMLR, 2020.

\bibitem[Lin et~al.(2025)Lin, Jin, and Jordan]{lin2025two}
Tianyi Lin, Chi Jin, and Michael~I Jordan.
\newblock Two-timescale gradient descent ascent algorithms for nonconvex minimax optimization.
\newblock \emph{Journal of Machine Learning Research}, 26\penalty0 (11):\penalty0 1--45, 2025.

\bibitem[Lu(2022)]{lu2022sr}
Haihao Lu.
\newblock An o (sr)-resolution ode framework for understanding discrete-time algorithms and applications to the linear convergence of minimax problems.
\newblock \emph{Mathematical Programming}, 194\penalty0 (1):\penalty0 1061--1112, 2022.

\bibitem[Mazumdar et~al.(2020)Mazumdar, Ratliff, and Sastry]{mazumdar2020gradient}
Eric Mazumdar, Lillian~J Ratliff, and S~Shankar Sastry.
\newblock On gradient-based learning in continuous games.
\newblock \emph{SIAM Journal on Mathematics of Data Science}, 2\penalty0 (1):\penalty0 103--131, 2020.

\bibitem[Mazumdar et~al.(2025)Mazumdar, Sastry, and Jordan]{mazumdar2019finding}
Eric Mazumdar, S~Shankar Sastry, and Michael~I Jordan.
\newblock On finding local nash equilibria (and only local nash equilibria) in zero-sum games.
\newblock \emph{ACM/IMS Journal of Data Science}, 2\penalty0 (2):\penalty0 1--26, 2025.

\bibitem[McKelvey and Palfrey(1998)]{mckelvey1998quantal}
Richard~D McKelvey and Thomas~R Palfrey.
\newblock Quantal response equilibria for extensive form games.
\newblock \emph{Experimental economics}, 1:\penalty0 9--41, 1998.

\bibitem[Mertikopoulos et~al.(2018)Mertikopoulos, Papadimitriou, and Piliouras]{mertikopoulos2018cycles}
Panayotis Mertikopoulos, Christos Papadimitriou, and Georgios Piliouras.
\newblock Cycles in adversarial regularized learning.
\newblock In \emph{Proceedings of the twenty-ninth annual ACM-SIAM symposium on discrete algorithms}, pages 2703--2717. SIAM, 2018.

\bibitem[Mertikopoulos et~al.(2019)Mertikopoulos, Lecouat, Zenati, Foo, Chandrasekhar, and Piliouras]{mertikopoulos2018optimistic}
Panayotis Mertikopoulos, Bruno Lecouat, Houssam Zenati, Chuan-Sheng Foo, Vijay Chandrasekhar, and Georgios Piliouras.
\newblock Optimistic mirror descent in saddle-point problems: Going the extra (gradient) mile.
\newblock In \emph{Proceedings of the International Conference on Learning Representations (ICLR)}, 2019.

\bibitem[Mescheder et~al.(2017)Mescheder, Nowozin, and Geiger]{mescheder2017numerics}
Lars Mescheder, Sebastian Nowozin, and Andreas Geiger.
\newblock The numerics of gans.
\newblock \emph{Advances in neural information processing systems}, 30, 2017.

\bibitem[Mokhtari et~al.(2020)Mokhtari, Ozdaglar, and Pattathil]{mokhtari2020unified}
Aryan Mokhtari, Asuman Ozdaglar, and Sarath Pattathil.
\newblock A unified analysis of extra-gradient and optimistic gradient methods for saddle point problems: Proximal point approach.
\newblock In \emph{International Conference on Artificial Intelligence and Statistics}, pages 1497--1507. PMLR, 2020.

\bibitem[Nocedal and Wright(1999)]{nocedal1999numerical}
Jorge Nocedal and Stephen~J Wright.
\newblock \emph{Numerical Optimization}.
\newblock Springer, 1999.

\bibitem[Pearlmutter(1994)]{pearlmutter1994fast}
Barak~A Pearlmutter.
\newblock Fast exact multiplication by the hessian.
\newblock \emph{Neural computation}, 6\penalty0 (1):\penalty0 147--160, 1994.

\bibitem[Ratliff et~al.(2016)Ratliff, Burden, and Sastry]{ratliff2016characterization}
Lillian~J Ratliff, Samuel~A Burden, and S~Shankar Sastry.
\newblock On the characterization of local nash equilibria in continuous games.
\newblock \emph{IEEE transactions on automatic control}, 61\penalty0 (8):\penalty0 2301--2307, 2016.

\bibitem[Rockafellar(1976)]{rockafellar1976monotone}
R~Tyrrell Rockafellar.
\newblock Monotone operators and the proximal point algorithm.
\newblock \emph{SIAM journal on control and optimization}, 14\penalty0 (5):\penalty0 877--898, 1976.

\bibitem[Roth(2002)]{roth2002economist}
Alvin~E Roth.
\newblock The economist as engineer: Game theory, experimentation, and computation as tools for design economics.
\newblock \emph{Econometrica}, 70\penalty0 (4):\penalty0 1341--1378, 2002.

\bibitem[Roughgarden(2010)]{roughgarden2010algorithmic}
Tim Roughgarden.
\newblock Algorithmic game theory.
\newblock \emph{Communications of the ACM}, 53\penalty0 (7):\penalty0 78--86, 2010.

\bibitem[Rubinstein(1982)]{rubinstein1982perfect}
Ariel Rubinstein.
\newblock Perfect equilibrium in a bargaining model.
\newblock \emph{Econometrica: Journal of the Econometric Society}, pages 97--109, 1982.

\bibitem[Sastry(1999)]{sastry2013nonlinear}
Shankar Sastry.
\newblock \emph{Nonlinear systems: analysis, stability, and control}.
\newblock Springer New York, 1999.

\bibitem[Sion(1958)]{sion1958general}
Maurice Sion.
\newblock On general minimax theorems.
\newblock \emph{Pacific Journal of Mathematics}, 8\penalty0 (1):\penalty0 171--176, 1958.

\bibitem[Thekumparampil et~al.(2019)Thekumparampil, Jain, Netrapalli, and Oh]{thekumparampil2019efficient}
Kiran~K Thekumparampil, Prateek Jain, Praneeth Netrapalli, and Sewoong Oh.
\newblock Efficient algorithms for smooth minimax optimization.
\newblock \emph{Advances in neural information processing systems}, 32, 2019.

\bibitem[Tseng(1995)]{tseng1995linear}
Paul Tseng.
\newblock On linear convergence of iterative methods for the variational inequality problem.
\newblock \emph{Journal of Computational and Applied Mathematics}, 60\penalty0 (1-2):\penalty0 237--252, 1995.

\bibitem[von Neumann(1928)]{v1928theorie}
J~von Neumann.
\newblock Zur theorie der gesellschaftsspiele.
\newblock \emph{Mathematische annalen}, 100\penalty0 (1):\penalty0 295--320, 1928.

\bibitem[Wang et~al.(2020)Wang, Zhang, and Ba]{Wang*2020On}
Yuanhao Wang, Guodong Zhang, and Jimmy Ba.
\newblock On solving minimax optimization locally: A follow-the-ridge approach.
\newblock In \emph{International Conference on Learning Representations}, 2020.
\newblock URL \url{https://openreview.net/forum?id=Hkx7_1rKwS}.

\end{thebibliography}
